\documentclass[amsmath,amssymb,aps,10pt,prd,twocolumn,letterpaper,nofootinbib,balancelastpage,notitlepage,superscriptaddress,floatfix,preprintnumbers]{revtex4-2}
\usepackage{graphicx}	% Standard figures 
\usepackage{ragged2e}	% for justified text in JHEP.bst
\usepackage{bm}		    % Bold math
\usepackage{empheq}     % for boxes in align environment
\newcommand*\widefbox[1]{\fbox{\hspace{1em}#1\hspace{1em}}}
\usepackage[sort&compress]{natbib}	 	% Standard reference package
	\setcitestyle{square,numbers,comma}	% Set citation style
\usepackage[colorlinks=true,urlcolor=blue,linkcolor=red,citecolor=red]{hyperref}	

\usepackage{mwe}
\usepackage{graphicx}
\usepackage{subcaption}
\usepackage{caption}
\usepackage{soul}

\newcommand{\tabref}[2][]{Table{#1}~\ref{tab:#2}}		% Table reference
\newcommand{\figref}[2][]{Fig{#1}.~\ref{fig:#2}}		% Figure reference
\newcommand{\secref}[2][]{Sec{#1}.~\ref{sec:#2}}		% Section reference
\newcommand{\appref}[2][x]{Appendi{#1}~\ref{app:#2}}	% Appendix reference
\renewcommand{\eqref}[2][]{Eq{#1}.~(\ref{eq:#2})}		% Equation reference
\newcommand{\eqrefRange}[2]{Eqs.~(\ref{eq:#1})--(\ref{eq:#2})}		% Equation range reference
\newcommand{\citeR}[2][]{Ref{#1}.~\cite{#2}}			% Ref. Citation
\newcommand{\nl}{\nonumber \\ & \quad }					% in-array new line
\newcommand{\Hz}[1][]{\,\mathrm{{#1}Hz}}
\newcommand{\s}[1][]{\,\mathrm{{#1}s}}
\DeclareMathOperator{\RE}{\mathrm{Re}}
\DeclareMathOperator{\IM}{\mathrm{Im}}
\begin{document}
%%%%%%%%%%%%%%%%%%%%%%%%%%%%%%%%%%%%%%%%%%%%%%%%%%%%%%%%%%%%%%%%%%%%%%%%%%%%%%%%%%%%%%%%%%
%%%%%%%%%%%%%%%%%%%%%%%%%%%%%%%%%%%%%%%%%%%%%%%%%%%%%%%%%%%%%%%%%%%%%%%%%%%%%%%%%%%%%%%%%%
%%%%%%%%%%%%%%%%%%%%%%%%%%%%%%%%%%%%%%%%%%%%%%%%%%%%%%%%%%%%%%%%%%%%%%%%%%%%%%%%%%%%%%%%%%
%%%%%%%%%%%%%%%%%%%%%%%%%%%%%%%%%%%%%%%%%%%%%%%%%%%%%%%%%%%%%%%%%%%%%%%%%%%%%%%%%%%%%%%%%%

%%%%%%%%%%%%%%%%%%%%%%%%%%%%%%%%%%%%%%%%%%%%%%%%%%%%%%%%%%%%%%%%%%%%%%%%%%%%%%%%%%%%%%%%%%
% Title, Author and Affiliation
\title{Modeling frequency instability in high-quality resonant experiments}
\date{\today}
%%%%%%%%%%%%%%%%%%%%%%%%%%%%%%
\author{Hao-Ran Cui}
\email{cui00159@umn.edu}
\affiliation{School of Physics \& Astronomy, University of Minnesota, Minneapolis, MN 55455, USA}
%%%%%%%%%%%%%%%%%%%%%%%%%%%%%%
\author{Saarik Kalia}
\email{kalias@umn.edu}
\affiliation{School of Physics \& Astronomy, University of Minnesota, Minneapolis, MN 55455, USA}
%%%%%%%%%%%%%%%%%%%%%%%%%%%%%%
\author{Zhen Liu}
\email{zliuphys@umn.edu}
\affiliation{School of Physics \& Astronomy, University of Minnesota, Minneapolis, MN 55455, USA}
%%%%%%%%%%%%%%%%%%%%%%%%%%%%%%%%%%%%%%%%%%%%%%%%%%%%%%%%%%%%%%%%%%%%%%%%%%%%%%%%%%%%%%%%%%

%%%%%%%%%%%%%%%%%%%%%%%%%%%%%%%%%%%%%%%%%%%%%%%%%%%%%%%%%%%%%%%%%%%%%%%%%%%%%%%%%%%%%%%%%%
\preprint{UMN-TH-4420/25}
\preprint{FERMILAB-PUB-25-0260-SQMS}
%%%%%%%%%%%%%%%%%%%%%%%%%%%%%%%%%%%%%%%%%%%%%%%%%%%%%%%%%%%%%%%%%%%%%%%%%%%%%%%%%%%%%%%%%%

%%%%%%%%%%%%%%%%%%%%%%%%%%%%%%%%%%%%%%%%%%%%%%%%%%%%%%%%%%%%%%%%%%%%%%%%%%%%%%%%%%%%%%%%%%
% Abstract
\begin{abstract}
%%%%%%%%%%%%%%%%%%%%%%%%%%%%%
Modern resonant sensing tools can achieve increasingly high quality factors, which correspond to extremely narrow linewidths.  In such systems, time-variation of the resonator's natural frequency can potentially impact its ability to accumulate power and its resulting sensitivity.  One such example is the Dark SRF experiment, which utilizes superconducting radio frequency (SRF) cavities with quality factors of $Q\sim10^{10}$.  Microscopic deformations of the cavity lead to stochastic jittering of its resonant frequency with amplitude 20 times its linewidth.  Naively, one may expect this to lead to a large suppression in accumulated power.  In this work, we study in detail the effects of frequency instability on high-quality resonant systems, utilizing the Dark SRF experiment as a case study.  We show that the timescale of jittering is crucial to determining its effect on power accumulation.  Namely, when the resonant frequency varies sufficiently quickly, the system accumulates power as if there were no jittering at all.  This implies that the sensitivity of a jittering resonator is comparable to that of a stable resonator.  In the case of Dark SRF, we find that jittering only induces a $\sim 10\%$ loss in power.  Our results allow the dark-photon exclusion bound from Dark SRF's pathfinder run to be refined, leading to a constraint that is an order of magnitude stronger than previously reported (corresponding to a signal-to-noise ratio which is four orders of magnitude larger).  This result represents the world-leading constraint on dark photons over a wide range of masses below $6\,\rm \mu eV$ and translates to the best laboratory-based limits on the photon mass $m_\gamma<2.9\times 10^{-48}\,\rm g$.
%%%%%%%%%%%%%%%%%%%%%%%%%%%%%
\end{abstract}
%%%%%%%%%%%%%%%%%%%%%%%%%%%%%%%%%%%%%%%%%%%%%%%%%%%%%%%%%%%%%%%%%%%%%%%%%%%%%%%%%%%%%%%%%%

%%%%%%%%%%%%%%%%%%%%%%%%%%%%%%%%%%%%%%%%%%%%%%%%%%%%%%%%%%%%%%%%%%%%%%%%%%%%%%%%%%%%%%%%%%
\maketitle
%%%%%%%%%%%%%%%%%%%%%%%%%%%%%%%%%%%%%%%%%%%%%%%%%%%%%%%%%%%%%%%%%%%%%%%%%%%%%%%%%%%%%%%%%%

%%%%%%%%%%%%%%%%%%%%%%%%%%%%%%%%%%%%%%%%%%%%%%%%%%%%%%%%%%%%%%%%%%%%%%%%%%%%%%%%%%%%%%%%%%
%%%%%%%%%%%%%%%%%%%%%%%%%%%%%%%%%%%%%%%%%%%%%%%%%%%%%%%%%%%%%%%%%%%%%%%%%%%%%%%%%%%%%%%%%%
\section{Introduction}
\label{sec:introduction}
%%%%%%%%%%%%%%%%%%%%%%%%%%%%%%%%%%%%%%%%%%%%%%%%%%%%%%%%%%%%%%%%%%%%%%%%%%%%%%%%%%%%%%%%%%
%%%%%%%%%%%%%%%%%%%%%%%%%%%%%%%%%%%%%%%%%%%%%%%%%%%%%%%%%%%%%%%%%%%%%%%%%%%%%%%%%%%%%%%%%%

Precision sensing tools are utilized in a wide variety of applications, ranging from ultraprecise clocks~\cite{Ludlow2015,Peik2021} to measurements of magnetic~\cite{Budker2013,Kimball2016} or gravitational fields~\cite{Goodkind1999} to searches for new particles~\cite{Carney2021,Romanenko2023,Higgins2024}.  Many of these systems rely on the use of increasingly high-quality resonators~\cite{Romanenko2014,Vinante2021,Hofer2023}, which in certain contexts can achieve quality factors as high as $Q\sim10^{11}$.  As the linewidths of these resonators shrink, the effects of frequency instability become increasingly important, and accurate modeling of these effects is necessary in order to reliably determine the sensitivity of these systems.

A notable recent example is the Dark SRF experiment~\cite{Romanenko2023}, a cavity-based ``light-shining-through-walls" (LSW) search for dark photons~\cite{Okun:1982xi,Bibber1987,JAECKEL2008509,Graham2014}.  This experiment makes use of superconducting radio frequency (SRF) cavities, with quality factors of $Q\sim10^{10}$, to produce and detect dark photons, which can mix with the Standard Model (SM) photon.  An emitter cavity is driven at its resonant frequency $f_0\sim1.3\Hz[G]$.  The SM photons in this cavity can convert to dark photons of the same frequency, which can in turn excite a receiver cavity, whose resonant frequency is precisely matched to $f_0$.  As the linewidths of these resonant cavities are extremely narrow ($\sim0.1\Hz$), small deviations in the resonant frequency of the cavity can disrupt the resonant enhancement of the experiment.  Other future cavity-based experiments are also subject to and have analyzed similar deviations in the resonant frequency~\cite{Berlin_2021}.

In \citeR{Romanenko2023}, two such deviations were discussed: a slow secular drift occurring on longer timescales of minutes; and a fast jittering, known as ``microphonics", which occurs on shorter $\mathcal O(10)\,\mathrm{ms}$ timescales.  The latter is a stochastic effect, which can arise from nanometer-scale deformations of the cavity, e.g. due to bubble collisions from the cooling fluid.  \citeR{Romanenko2023} took a conservative approach in accounting for this frequency instability, modeling its effect as if the cavity frequencies were always mismatched.%
%%%%%%%%%%%%
\footnote{This conservative approach was adapted because the drifting and jittering are both $\sim\mathrm{few}\Hz$, and the readout is an averaged power spectrum from the experiment.  Given the similar size of these frequency-changing effects and a lack of proper modeling of the jittering, they were both taken as a constant frequency mismatch between the emitter and receiver cavities.}
%%%%%%%%%%%%
This led to a suppression of the estimated signal power and resulting signal-to-noise ratio (SNR) of $\sim10^{-5}$.  In this work, we more precisely model the effects of stochastic frequency instability and show that the suppression is not nearly as severe.

We primarily focus on two effects of jittering in resonant systems.  The first is its impact on power accumulation in the resonator.  A naive expectation is that jittering should suppress power accumulation if the amplitude of jittering is larger than the linewidth of the resonator, as it would cause the resonator to spend significant periods of time off-resonance.  We show that the timescale of the jittering is important to its effect on power suppression.  In particular, if the resonant frequency jitters sufficiently quickly, the system can actually accumulate power as if there were no jittering at all!  This is because power suppression occurs when the resonator develops a relative phase with the driving force.  When the system spends a significant amount of time at a fixed off-resonant frequency, a large phase can develop.  However, if the resonant frequency is jittering back and forth quickly about the driving frequency, then a relative phase will not develop.

The second effect is the impact of jittering on the spectral response of the system, and in turn, its sensitivity.  In this work, we show that jittering introduces richer spectral structure into the response of the resonator.  In particular, it can lead to sidebands corresponding to the dominant frequencies of jittering.  Meanwhile, jittering does not significantly affect the central resonant peak in the spectral response, when its impact on power accumulation is small.  As the system's sensitivity is dominated by its response on-resonance, this implies that jittering does not degrade the sensitivity of a resonant system.

This work is organized as follows.  In \secref{resonator}, we introduce a toy model of a driven resonator with stochastic frequency variations and outline how its dynamics can be numerically solved.  In \secref{characteristics}, we compute various characteristics of the system, including its expected amplitude, expected power, and correlation function.  We define a perturbative regime where the power suppression is small and analytic results can be computed, and we show that the Dark SRF experiment lies in this regime.  In \secref{spectral}, we calculate the spectral response of a jittering resonator and use it to compute the sensitivity of the system to a narrowband signal.  Finally, in \secref{discussion}, we discuss the implications of this work for future experiments, including the re-interpretation of Dark SRF's existing data.  In \appref[ces]{perturbative} and \ref{app:shapiro}, we perform analytic calculations of various quantities relevant to the jittering resonator system.  In \appref{SNR}, we derive the figure of merit for sensitivity used in this work.  We make all the code used in this work publicly available on Github~\cite{github}.

%%%%%%%%%%%%%%%%%%%%%%%%%%%%%%%%%%%%%%%%%%%%%%%%%%%%%%%%%%%%%%%%%%%%%%%%%%%%%%%%%%%%%%%%%%
%%%%%%%%%%%%%%%%%%%%%%%%%%%%%%%%%%%%%%%%%%%%%%%%%%%%%%%%%%%%%%%%%%%%%%%%%%%%%%%%%%%%%%%%%%
\section{Jittering resonator}
\label{sec:resonator}
%%%%%%%%%%%%%%%%%%%%%%%%%%%%%%%%%%%%%%%%%%%%%%%%%%%%%%%%%%%%%%%%%%%%%%%%%%%%%%%%%%%%%%%%%%
%%%%%%%%%%%%%%%%%%%%%%%%%%%%%%%%%%%%%%%%%%%%%%%%%%%%%%%%%%%%%%%%%%%%%%%%%%%%%%%%%%%%%%%%%%

We consider a resonator described by $x(t)$, which satisfies the differential equation%
%%%%%%%%%%%%
\footnote{In cases where $x(t)$ and $F(t)$ represent a physical position and physical force, the right-hand side (RHS) of \eqref{resonator} should be $F(t)/m$.  For notational convenience, in this work, we set the mass $m=1$.}
%%%%%%%%%%%%
\begin{equation}
    \ddot x(t) + \gamma \dot x(t) + (\omega_0 +\delta \omega (t))^2 x(t) = F(t).
    \label{eq:resonator}
\end{equation}
Here $\omega_0+\delta\omega(t)$ is the real-time natural frequency of the resonator,%
%%%%%%%%%%%%
\footnote{Throughout this work, we will refer to both the ordinary frequency $f_\mathrm{label}$ and angular frequency $\omega_\mathrm{label}$ for various quantities.  These are always related by $\omega_\mathrm{label}=2\pi f_\mathrm{label}$.}
%%%%%%%%%%%%
with the jittering effect represented by $\delta\omega(t)$; $\gamma=\omega_0/Q$ is the linewidth of the resonator;%
%%%%%%%%%%%%
\footnote{In this work, we take $\gamma=\gamma_\mathrm{loaded}$ to include dissipation from sources external to the cavity.  For the cavity described in \tabref{parameters}, the linewidth associated only with the cavity is $\gamma_\mathrm{cavity}\approx2\pi\times0.04\Hz$~\cite{Romanenko2023}.}
%%%%%%%%%%%%
and $F(t)$ represents a driving force.  In this section and \secref{characteristics}, we will consider the driving force to be monochromatic%
%%%%%%%%%%%%
\footnote{In many contexts it is more appropriate for the force $F_R(t)=F_0\cos(\omega_F t)$ to be a real quantity.  In this work, we will deal with the complexified version $F_C(t)$ in \eqref{force}.  The response $x_R(t)$ to a real force $F_R(t)$ can always be found by solving for the response $x_C(t)$ to a complex force $F_C(t)$, and taking the real part $x_R(t)=\RE[x_C(t)]$.}
%%%%%%%%%%%%
\begin{equation}
    F(t)=F_0e^{i\omega_Ft},
    \label{eq:force}
\end{equation}
with frequency $\omega_F=\omega_0+\Delta\omega_F$ close to the natural frequency of the resonator.%
%%%%%%%%%%%%
\footnote{In this work, all effects of ``drifting" will be represented by $\Delta\omega_F$.  That is, any variations which are longer than the timescale over which the average $\langle\cdot\rangle_\infty$ is taken are treated as a constant frequency offset.  Variations over shorter timescales will be included in the low-frequency part of the jittering spectrum $S_{\delta\omega}(\omega)$.}
%%%%%%%%%%%%
In the context of a mechanical resonator, $x(t)$ might represent the position of the resonator, while $F(t)$ represents a real force applied to the system.  In the context of a resonant cavity, such as Dark SRF, $x(t)$ represents the electric field amplitude of a resonant mode, while $F(t)$ represents currents which excite the resonant mode.  (These could be either physical currents which generate thermal noise, or an effective current sourced by a dark-photon field~\cite{Graham2014}.)  Regardless of context, we will refer to $x(t)$ as the \emph{position} of the resonator and $F(t)$ as a \emph{force}.

In this work, we will treat $\delta\omega(t)$ as a random process, so that \eqref{resonator} is a stochastic differential equation.  We will therefore be interested in computing ensemble-averaged quantities. In addition, \eqref{resonator} represents a driven damped harmonic oscillator, so the system will exhibit transient behavior that depends on its initial conditions, but its asymptotic behavior will not.  In \secref{characteristics}, we will compute quantities in the $t\rightarrow\infty$ limit and averaged over different realizations of $\delta\omega(t)$.  We will denote such asymptotic ensemble-averaged quantities by $\langle\cdot\rangle_\infty$.  Quantities of interest will include the mean amplitude $\langle x(t)\rangle_\infty$ and the mean power $\langle|x(t)|^2\rangle_\infty$.

In the ``no-jittering case" where $\delta\omega(t)=0$ [denoted by $x_0(t)$], the familiar asymptotic results for these quantities are
\begin{align}
    \label{eq:nojittering_amp}
    \lim_{t\rightarrow\infty}x_0(t)&=\frac{F_0e^{i\omega_Ft}}{i\gamma\omega_F+\omega_0^2-\omega_F^2},\\
    \lim_{t\rightarrow\infty}|x_0(t)|^2&=\frac{|F_0|^2}{\gamma^2\omega_F^2+(\omega_0^2-\omega_F^2)^2}.
    \label{eq:nojittering_power}
\end{align}
Meanwhile, for the ``fixed off-resonance case" where $\delta\omega(t)=\delta\omega_0$ [denoted by $x_\mathrm{fix}(t)$], these are given by
\begin{align}
    \lim_{t\rightarrow\infty}x_\mathrm{fix}(t)&=\frac{F_0e^{i\omega_Ft}}{i\gamma\omega_F+(\omega_0+\delta\omega_0)^2-\omega_F^2},\\
    \lim_{t\rightarrow\infty}|x_\mathrm{fix}(t)|^2&=\frac{|F_0|^2}{\gamma^2\omega_F^2+((\omega_0+\delta\omega_0)^2-\omega_F^2)^2}.
\end{align}
Ultimately, we will be interested in studying the case of a very good resonator, that is, when the central natural frequency is much larger than all other quantities $\omega_0\gg\gamma,\delta\omega_0,\Delta\omega_F$.  In this limit,%
%%%%%%%%%%%%
\footnote{As $|x_0(t)|^2$ and $|x_\mathrm{fix}(t)|^2$ are independent of $t$ in the $t\rightarrow\infty$ limit, we will use $|x_0(\infty)|^2$ and $|x_\mathrm{fix}(\infty)|^2$ as shorthands for this limit.}
%%%%%%%%%%%%
\begin{align}
    |x_\mathrm{fix}(\infty)|^2&=\frac{|F_0|^2}{\omega_0^2\left(\gamma^2+4(\delta\omega_0-\Delta\omega_F)^2\right)}\\
    &=\frac{\gamma^2}{\gamma^2+4(\delta\omega_0-\Delta\omega_F)^2}\cdot\left[|x_0(\infty)|^2\right]_{\Delta\omega_F=0}.
    \label{eq:suppression}
\end{align}

%%%%%%%%%%%%%%%%%%%%%%%%%%%%%%%%%%%%%%%%%%
%%%%%%%%%%%%%%%%%%%%%%%%%%%%%%%%%%%%%%%%%%
\begin{figure}[h!]
    \centering
    \includegraphics[width=0.95\columnwidth]{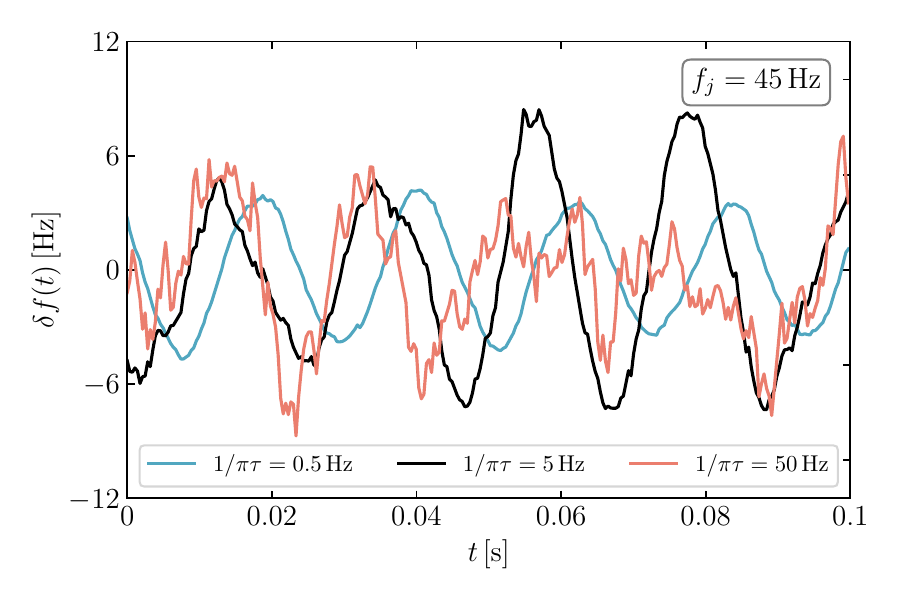}
    \includegraphics[width=0.95\columnwidth]{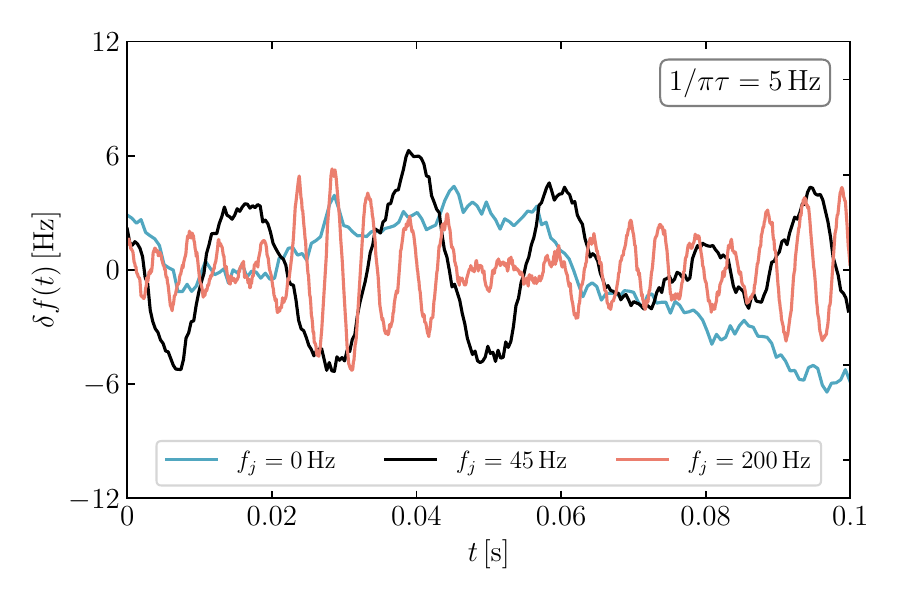}
    \includegraphics[width=0.95\columnwidth]{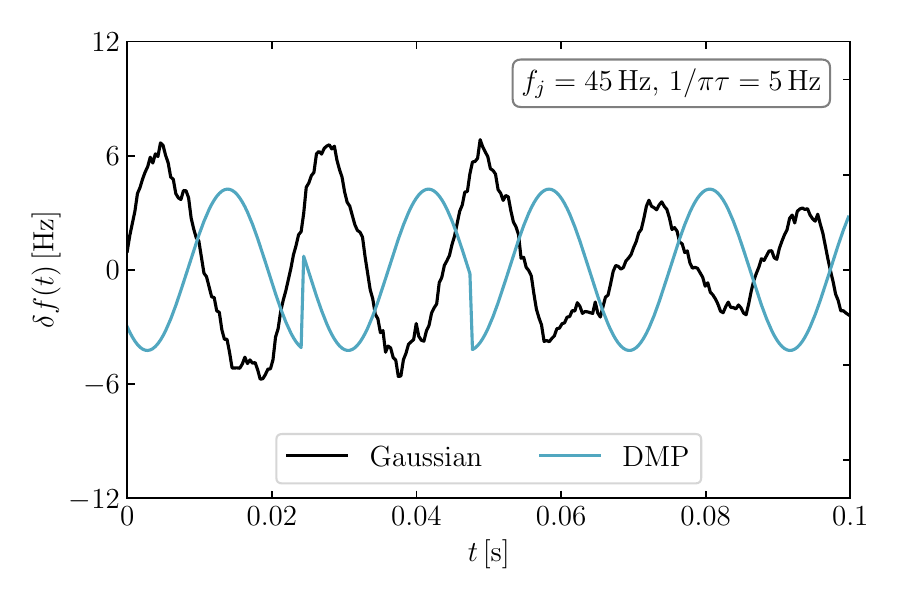}
    \caption{Realizations of $\delta f(t)=\delta\omega(t)/2\pi$ for various choices of parameters.  In each plot, the black curve utilizes the benchmark parameters for Dark SRF shown in \tabref{parameters}, and the colored curves show the behavior when a single parameter is varied.  The upper plot shows the dependence on the correlation time $\tau$, the middle plot shows the dependence on the peak jittering frequency $f_j$, and the lower plot compares the Gaussian and DMP models of jittering.}
    \label{fig:jittering}
\end{figure}
%%%%%%%%%%%%%%%%%%%%%%%%%%%%%%%%%%%%%%%%%%
%%%%%%%%%%%%%%%%%%%%%%%%%%%%%%%%%%%%%%%%%%

The prefactor in \eqref{suppression} represents the power suppression resulting from a fixed frequency mismatch between the driving force and the resonator.  This is the worst-case power suppression that could occur from a jittering effect of typical amplitude $\sim\delta\omega_0$.  For instance, \citeR{Romanenko2023} utilized this prefactor in their analysis in order to take a conservative approach to modeling the effect of jittering and drifting.  In \secref{characteristics}, we will show that the true power suppression from jittering may be far less severe than this.  In fact, if the jittering occurs on a fast enough timescale, there may be barely any suppression at all! 

%%%%%%%%%%%%%%%%%%%%%%%%%%%%%%%%%%%%%%%%%%%%%%%%%%%%%%%%%%%%%%%%%%%%%%%%%%%%%%%%%%%%%%%%%%
%%%%%%%%%%%%%%%%%%%%%%%%%%%%%%%%%%%%%%%%%%%%%%%%%%%%%%%%%%%%%%%%%%%%%%%%%%%%%%%%%%%%%%%%%%
\subsection{Modeling $\delta\omega(t)$}
\label{sec:jittering}
%%%%%%%%%%%%%%%%%%%%%%%%%%%%%%%%%%%%%%%%%%%%%%%%%%%%%%%%%%%%%%%%%%%%%%%%%%%%%%%%%%%%%%%%%%
%%%%%%%%%%%%%%%%%%%%%%%%%%%%%%%%%%%%%%%%%%%%%%%%%%%%%%%%%%%%%%%%%%%%%%%%%%%%%%%%%%%%%%%%%%

In order to solve \eqref{resonator}, we must first define a model for the jittering $\delta\omega(t)$, which will be used both in our analytic and numerical calculations.  In this work, we will assume that $\delta\omega(t)$ is stationary and has zero mean $\langle\delta\omega(t)\rangle=0$ (since any nonzero mean can be absorbed into $\omega_0$). 
 The statistics of $\delta\omega(t)$ are then primarily characterized by its autocorrelation function, or equivalently, by its power spectral density (PSD)
\begin{align}
    \label{eq:jittering_corr}
    C_{\delta\omega}(t-t')&\equiv\langle\delta\omega(t)\delta\omega(t')\rangle,\\
    S_{\delta\omega}(\omega)&\equiv\int C_{\delta\omega}(t) e^{-i\omega t}dt.
    \label{eq:jittering_PSD}
\end{align}

For simplicity, we will consider a PSD which exhibits a single peak for $\omega>0$.  (As $C_{\delta\omega}(t)$ and $S_{\delta\omega}(\omega)$ are both real and even, it must exhibit a reflected peak for $\omega<0$.)  Numerically, it will be simplest to simulate a peak described by a Lorentzian.  Therefore in our work, we fix
\begin{equation}
    S_{\delta\omega}(\omega)=\frac{\delta\omega_0^2}{\tau}\left(\frac1{\tau^{-2}+(\omega-\omega_j)^2}+\frac1{\tau^{-2}+(\omega+\omega_j)^2}\right),
    \label{eq:Sdelomega}
\end{equation}
where $\delta\omega_0$ represents the typical amplitude of jittering, $\omega_j$ is its peak frequency, and $2/\tau$ is the linewidth of the peak.  (In \appref{perturbative}, we also include results for general $S_{\delta\omega}$.)  Benchmark parameter values for the receiver cavity in the Dark SRF experiment are displayed in \tabref{parameters}.  The PSD in \eqref{Sdelomega} corresponds to an autocorrelation function
\begin{align}
    C_{\delta\omega}(t)&=\delta\omega_0^2 \cos (\omega_jt)e^{-|t|/\tau}.
    \label{eq:correlation}
\end{align}
We see that the dominant jittering frequency $\omega_j$ gives an oscillatory pattern to $\delta\omega(t)$, and $\tau$ represents the timescale on which its correlations decay.  The upper and middle plots of \figref{jittering} show how the behavior of $\delta\omega(t)$ varies with $\tau$ and $\omega_j$, respectively.  Note that if $\omega_j\gg2/\tau$, the jittering is nearly oscillatory, while if $\omega_j\ll2/\tau$, the oscillatory behavior is washed out.

The cosine factor in \eqref{correlation} makes direct numerical construction of $\delta\omega(t)$ challenging. Therefore, we introduce the following decomposition
\begin{equation}
    \delta\omega(t)=\eta(t)\cos (\omega_j t)+\phi(t)\sin (\omega_j t),
    \label{eq:decomposition}
\end{equation}
where $\eta(t)$ and $\phi(t)$ are two independent random processes without the cosine factor in their autocorrelation functions,
\begin{align}
    \label{eq:etaphi1}
    &\langle \eta(t)\rangle=\langle \phi(t)\rangle=0,\\
    &\langle \eta(t+s)\eta(t)\rangle=\langle \phi(t+s)\phi(t)\rangle=\delta\omega_0^2 e^{-|s|/\tau},\label{eq:etaphi2}\\
    &\langle \eta(t+s)\phi(t)\rangle\equiv 0.
    \label{eq:etaphi3}
\end{align}
The autocorrelation function for $\delta\omega(t)$ in \eqref{correlation} then follows automatically. 

%%%%%%%%%%%%%%%%%%%%%%%%%%%%%%%%%%%%%%%%%%
%%%%%%%%%%%%%%%%%%%%%%%%%%%%%%%%%%%%%%%%%%
\begin{table}
    \centering
    \begin{tabular}{c c c}
        \hline\hline
        Parameter&Symbol&Value\\
        \hline
        Central natural frequency&$f_0=\frac{\omega_0}{2\pi}$&$1.3\,\mathrm{GHz}$\\
        Resonator linewidth&$\gamma$&$2\pi\times0.15\Hz$\\
        Jittering amplitude&$\delta f_0=\frac{\delta\omega_0}{2\pi}$&$3\Hz$\\
        Jittering correlation time&$\tau$&$\frac1{5\pi}\s$\\
        Peak jittering frequency&$f_j=\frac{\omega_j}{2\pi}$&$45\Hz$\\
        \hline\hline
    \end{tabular}
    \caption{Benchmark parameters for the Dark SRF experiment, based on \citeR[s]{Romanenko2023,pischalnikov2019operation}.  We model the receiver cavity as a jittering resonator, which satisfies \eqref{resonator}.  We model the jittering as a random process whose PSD has a single Lorentzian peak, as in \eqref{Sdelomega}.}
    \label{tab:parameters}
\end{table}
%%%%%%%%%%%%%%%%%%%%%%%%%%%%%%%%%%%%%%%%%%
%%%%%%%%%%%%%%%%%%%%%%%%%%%%%%%%%%%%%%%%%%

So far, we have specified the mean and autocorrelation functions of $\eta(t)$ and $\phi(t)$.  To fully define the statistics of the jittering, we must also fix the higher moments of these random processes.  In this work, we will consider two distinct models for $\eta(t)$ and $\phi(t)$, which fix these higher moments: a Gaussian process and a dichotomic Markov process (DMP).%
%%%%%%%%%%%%
\footnote{For a Gaussian process, any higher-point correlators $\langle\eta(t_1)\eta(t_2)\cdots\eta(t_n)\rangle$ can be decomposed into a sum of autocorrelation functions via Wick's theorem.  See \eqref{DMP_npoint} for the decomposition of higher point correlators for a DMP.}
%%%%%%%%%%%%

A Gaussian process $\eta(t)$ is defined so that for any set of discrete times $t_1,\ldots,t_n$, the random variables $\eta(t_1),\ldots,\eta(t_n)$ form a multivariate normal distribution with mean vector and covariance matrix given by \eqrefRange{etaphi1}{etaphi3}.  In particular, this implies that, at any fixed time $t_1$, the quantities $\eta(t_1)$ and $\phi(t_1)$ are Gaussian variables with mean zero and variance $\delta\omega_0^2$.  When utilizing Gaussian processes in our numerical simulations, we construct $\eta(t)$ and $\phi(t)$ by discretizing time and applying the method described in \citeR{deserno2002generate}.

A DMP is a Markov process that alternates between the discrete values $\pm\delta\omega_0$, where the time between alternations is described by an exponential random variable with expected value $\tau$.  A DMP $\eta(t)$ can be explicitly constructed by discretizing time into $t_k=k\Delta t$, randomly choosing $\eta(t_1)=\pm\delta\omega_0$ (with equal probability), and then setting $\eta(t_{k+1})=\eta(t_k)$ with probability $\frac{1+e^{-\Delta t/\tau}}2$ and $\eta(t_{k+1})=-\eta(t_k)$ otherwise, for all $k$.

It is natural to expect that physical jittering would be described by a Gaussian process.  For this reason, in most of our numerical simulations, we will model $\eta(t)$ and $\phi(t)$ as Gaussian processes.  Modeling them as DMPs, on the other hand, can make certain analytic calculations tractable~\cite{bourret1973brownian}.  We derive analytic results for the DMP case in \appref{shapiro}.  The lower plot of \figref{jittering} shows numerical simulations of $\delta\omega(t)$ for these two models, and \figref{GPDMPcomparison} compares the resulting power $|x(t)|^2$ in the resonator.  In general, we find that when the power suppression due to jittering is small, the higher moments of the random processes lead to higher-order corrections, and so the leading-order results for Gaussian processes and DMPs are similar.

%%%%%%%%%%%%%%%%%%%%%%%%%%%%%%%%%%%%%%%%%%%%%%%%%%%%%%%%%%%%%%%%%%%%%%%%%%%%%%%%%%%%%%%%%%
%%%%%%%%%%%%%%%%%%%%%%%%%%%%%%%%%%%%%%%%%%%%%%%%%%%%%%%%%%%%%%%%%%%%%%%%%%%%%%%%%%%%%%%%%%
\subsection{Numerical simulation}
\label{sec:numerics}
%%%%%%%%%%%%%%%%%%%%%%%%%%%%%%%%%%%%%%%%%%%%%%%%%%%%%%%%%%%%%%%%%%%%%%%%%%%%%%%%%%%%%%%%%%
%%%%%%%%%%%%%%%%%%%%%%%%%%%%%%%%%%%%%%%%%%%%%%%%%%%%%%%%%%%%%%%%%%%%%%%%%%%%%%%%%%%%%%%%%%

Here we outline how \eqref{resonator} can be solved numerically in a computationally efficient manner.  As we are interested in the case where $\omega_0$ is large, direct numerical evolution of \eqref{resonator} would require very precise time resolution.  However, if $\omega_j,1/\tau\ll\omega_0$, the relevant jittering dynamics occur on much longer timescales.  It is therefore beneficial to discretize our simulation with a much coarser resolution $1/\omega_0\ll\Delta t\ll1/\omega_j,\tau$.  Over the timescale $\Delta t$, the system will undergo many oscillations, but $\delta\omega(t)$ will be roughly constant, i.e. the system behaves as in the fixed off-resonance case.  \eqref{resonator} can therefore be solved analytically over every such interval.

More specifically, we discretize the total simulation time into $t_k=k\Delta t$.  For each simulation, we generate one realization of $\delta\omega(t)$ at all $t=t_k$ via the methods described in \secref{jittering}.  We also fix initial conditions $x(t_0)=\dot x(t_0)=0$.  The value of $x$ and $\dot x$ at the end of every interval can then be computed as
\begin{widetext}
\begin{align}
    \label{eq:xevol}
    x(t_{k+1})&=e^{-\frac{\gamma\Delta t}2}\left(x(t_k)\cos((\omega_0+\delta\omega(t_k))\Delta t)+\frac{\dot x(t_k)}{\omega_0}\sin((\omega_0+\delta\omega(t_k))\Delta t)\right)\nl
    +\frac{F_0e^{i\omega_F(t_k+\Delta t)}}{\omega_0(i\gamma+2\delta\omega(t_k)-2\Delta\omega_F)}\left(1-e^{-\frac{\gamma\Delta t}2+i(\delta\omega(t_k)-\Delta\omega_F)\Delta t}\right)\\
    \dot x(t_{k+1})&=e^{-\frac{\gamma\Delta t}2}\left(-\omega_0x(t_k)\sin((\omega_0+\delta\omega(t_k))\Delta t)+\dot x(t_k)\cos((\omega_0+\delta\omega(t_k))\Delta t)\right)\nl
    +\frac{iF_0e^{i\omega_F(t_k+\Delta t)}}{i\gamma+2\delta\omega(t_k)-2\Delta\omega_F}\left(1-e^{-\frac{\gamma\Delta t}2+i(\delta\omega(t_k)-\Delta\omega_F)\Delta t}\right).
    \label{eq:xdotevol}
\end{align}
\end{widetext}
As we are interested in asymptotic quantities, we should evolve until a time $t_n\equiv T\gg2/\gamma$ after which transient contributions have decayed.  In this way, we can compute $x(t)$ for a single realization of the jittering.  \figref{GPDMPcomparison} shows two such realizations, one with Gaussian jittering and one with DMP jittering.

The above procedure can provide us with a single realization of any quantity of interest $\mathcal O[x(t)]$.  In order to determine ensemble-averaged quantities, we must repeat this procedure to obtain several realizations of the system evolution $x_1(t),\ldots,x_N(t)$.  An asymptotic ensemble-averaged estimate for the mean and variance of $\mathcal O$ can then be obtained as
\begin{align}
    \label{eq:Omean}
    \langle\mathcal O\rangle_\infty&=\frac1N\sum_{i=1}^N\mathcal O[x_i(t)],\\
    \varsigma_\mathcal O^2&=\frac1{N-1}\sum_{i=1}^N\left(\mathcal O[x_i(t)]-\langle\mathcal O\rangle_\infty\right)^2.
    \label{eq:Ovar}
\end{align}
The statistical error on the estimate in \eqref{Omean} is simply $\varsigma_\mathcal O/\sqrt N$.  In our figures, we will also include a systematic error associated with transient effects due to the finite integration time $T$.  The total error on our estimate of $\langle\mathcal O\rangle_\infty$ is then given by
\begin{equation}
    \sigma_\mathcal O^2=\frac{\varsigma_\mathcal O^2}N+e^{-\gamma T}\langle\mathcal O\rangle_\infty,
\end{equation}
which we represent as a shaded band around our estimate.

%%%%%%%%%%%%%%%%%%%%%%%%%%%%%%%%%%%%%%%%%%%%%%%%%%%%%%%%%%%%%%%%%%%%%%%%%%%%%%%%%%%%%%%%%%
%%%%%%%%%%%%%%%%%%%%%%%%%%%%%%%%%%%%%%%%%%%%%%%%%%%%%%%%%%%%%%%%%%%%%%%%%%%%%%%%%%%%%%%%%%
\section{System characteristics}
\label{sec:characteristics}
%%%%%%%%%%%%%%%%%%%%%%%%%%%%%%%%%%%%%%%%%%%%%%%%%%%%%%%%%%%%%%%%%%%%%%%%%%%%%%%%%%%%%%%%%%
%%%%%%%%%%%%%%%%%%%%%%%%%%%%%%%%%%%%%%%%%%%%%%%%%%%%%%%%%%%%%%%%%%%%%%%%%%%%%%%%%%%%%%%%%%

In this section, we discuss various characteristics of the jittering resonator system.  First, we compute the expected power $\langle|x(t)|^2\rangle_\infty$ in the system, using the numerical procedure outlined in \secref{numerics}.  We also present analytic expressions for the expected power, which apply in the perturbative regime where the power suppression is small.  In particular, this allows us to define a perturbativity condition for when the impact of jittering on the signal power is negligible.  Next, we establish a connection between power accumulation and the phase mismatch between the resonator and driving force.  This allows us to understand why power suppression is small when the jittering is fast.  Finally, we study other statistics that characterize the system, such as the distribution of the signal power and temporal correlations.

%%%%%%%%%%%%%%%%%%%%%%%%%%%%%%%%%%%%%%%%%%%%%%%%%%%%%%%%%%%%%%%%%%%%%%%%%%%%%%%%%%%%%%%%%%
%%%%%%%%%%%%%%%%%%%%%%%%%%%%%%%%%%%%%%%%%%%%%%%%%%%%%%%%%%%%%%%%%%%%%%%%%%%%%%%%%%%%%%%%%%
\subsection{Expected signal power}
\label{sec:power}
%%%%%%%%%%%%%%%%%%%%%%%%%%%%%%%%%%%%%%%%%%%%%%%%%%%%%%%%%%%%%%%%%%%%%%%%%%%%%%%%%%%%%%%%%%
%%%%%%%%%%%%%%%%%%%%%%%%%%%%%%%%%%%%%%%%%%%%%%%%%%%%%%%%%%%%%%%%%%%%%%%%%%%%%%%%%%%%%%%%%%

We begin by solving \eqref{resonator} [with monochromatic $F(t)$ as in \eqref{force}] for the expected amplitude $\langle x(t)\rangle_\infty$ and power $\langle|x(t)|^2\rangle_\infty$ in the resonator.  In this subsection and \secref{distribution}, we fix $\omega_F=\omega_0$ and work in the limit $\omega_0\gg \gamma,\delta\omega_0,1/\tau,\omega_j$.  Various similar problems have been solved analytically in the literature.  For instance, \citeR[s]{bourret1973brownian,gitterman2005noisy} solved \eqref{resonator}, with $\delta\omega(t)$ modeled by a DMP and $\omega_j=0$, using a Green's function approach.  \citeR{burov2016noisy} utilized the Shapiro-Loginov formula~\cite{shapiro1978formulae} to solve a similar stochastic differential equation but with a random damping factor $\gamma(t)$ instead of natural frequency $\omega_0(t)$.  These works also primarily focused on the case where $F(t)$ is described by white noise. To our knowledge, there is no analytical solution in the literature which solves \eqref{resonator} with $\omega_j\neq0$ and/or Gaussian $\delta\omega(t)$.

%%%%%%%%%%%%%%%%%%%%%%%%%%%%%%%%%%%%%%%%%%
%%%%%%%%%%%%%%%%%%%%%%%%%%%%%%%%%%%%%%%%%%
\begin{figure}[t]
    \centering
    \includegraphics[width=0.95\columnwidth]{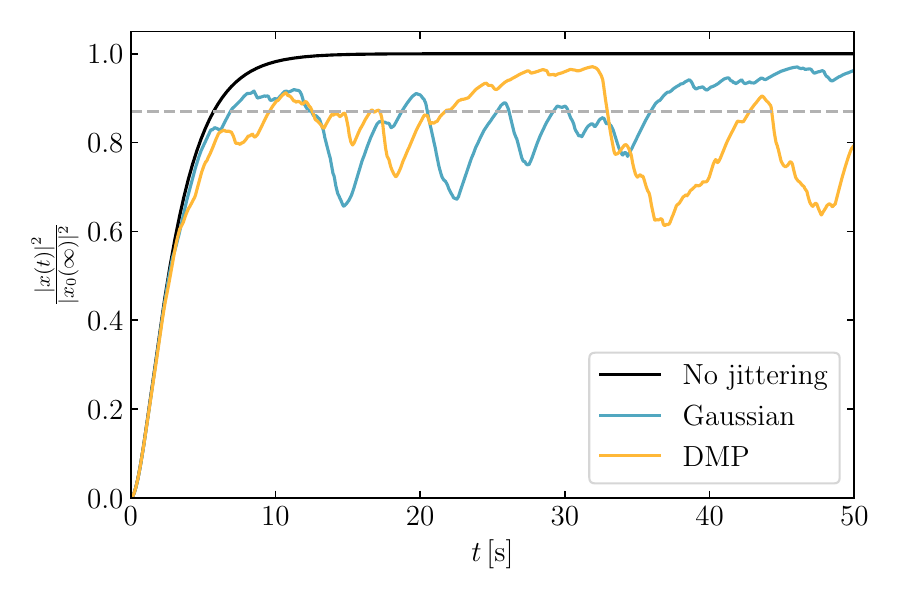}
    \caption{Real-time evolution of \eqref{resonator} for the parameters in \tabref{parameters} and both models of $\delta\omega(t)$.  We show the cases of Gaussian jittering (blue), DMP jittering (orange), and no jittering (black).  We plot the power $|x(t)|^2$ in the resonator, normalized by its asymptotic value in the no-jittering case [see \eqref{nojittering_power}].  The dashed grey line indicates the asymptotic ensemble-averaged power $\langle |x(\infty)|^2\rangle_\infty$ for the cases with nonzero jittering.  (The Gaussian and DMP cases give the same value of this quantity to within a percent.)}
    \label{fig:GPDMPcomparison}
\end{figure}
%%%%%%%%%%%%%%%%%%%%%%%%%%%%%%%%%%%%%%%%%%
%%%%%%%%%%%%%%%%%%%%%%%%%%%%%%%%%%%%%%%%%%

In \appref{perturbative}, we solve \eqref{resonator} in the perturbative regime, where the resonator amplitude can be approximated by the no-jittering case to leading order, $x(t)\approx x_0(t)$.  In this regime, only the two-point statistics of $\delta\omega(t)$ contribute to the leading order result, and so the results in this regime apply for both Gaussian and DMP jittering.  Specifically, we find
\begin{empheq}[box=\widefbox]{align}
    \label{eq:amplitude}
    \langle x(t)\rangle_\infty&\approx\frac{\lim_{t\rightarrow\infty}x_0(t)}{1+\alpha},\\
    \langle|x(t)|^2\rangle_\infty&\approx\frac{|x_0(\infty)|^2}{1+\alpha},
    \label{eq:power}
\end{empheq}
where the perturbative parameter $\alpha$ is given by
\begin{empheq}[box=\widefbox]{align}
    \label{eq:alpha}
    \alpha&\equiv\frac{4\delta\omega_0^2}{\gamma^2}\rho,\\
    \rho&\equiv\frac{\gamma\tau(2+\gamma\tau)}{(2+\gamma\tau)^2+4\omega_j^2\tau^2}.
    \label{eq:rho}
\end{empheq}
It is clear from \eqref[s]{amplitude} and (\ref{eq:power}) that when $\alpha\ll1$, then each instance of $x(t)$ should be close to $x_0(t)$.  Therefore, $\alpha$ is the appropriate parameter to define the perturbative regime, and when this parameter is small, power suppression due to jittering is negligible.  The Dark SRF parameters listed in \tabref{parameters} give $\alpha\approx0.15$.

%%%%%%%%%%%%%%%%%%%%%%%%%%%%%%%%%%%%%%%%%%
%%%%%%%%%%%%%%%%%%%%%%%%%%%%%%%%%%%%%%%%%%
\begin{figure}[h]
    \centering
    \includegraphics[width=0.95\columnwidth]{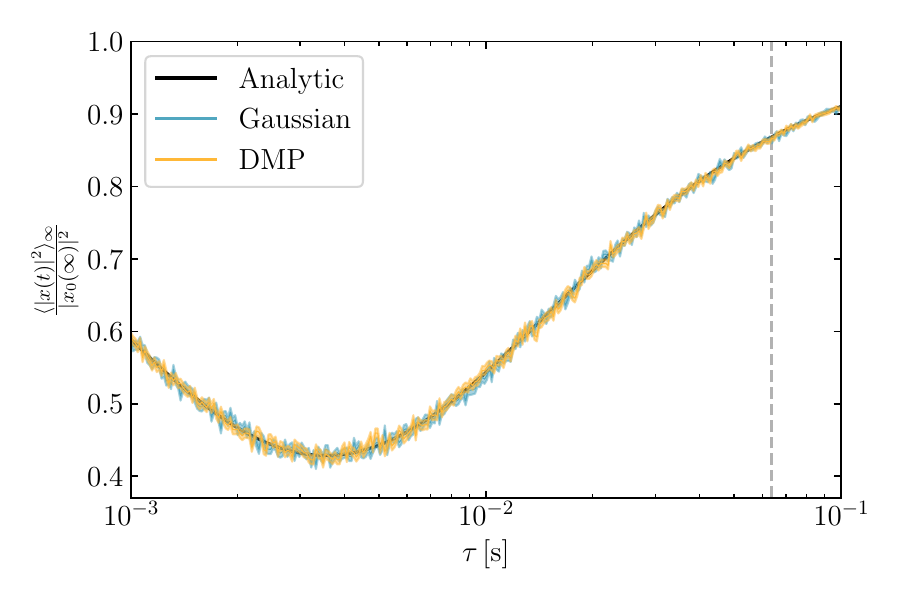}
    \includegraphics[width=0.95\columnwidth]{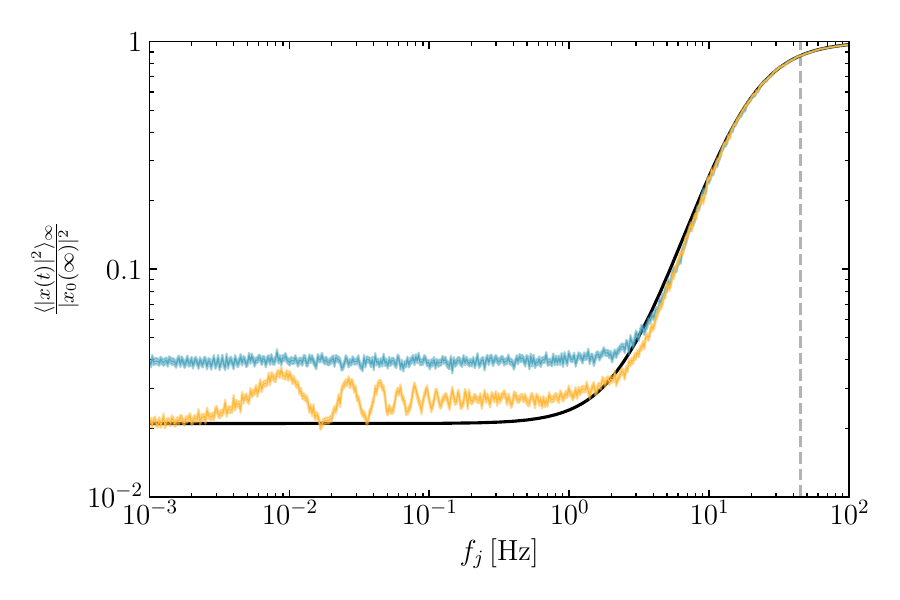}
    \caption{Dependence of expected power $\langle|x(t)|^2\rangle_\infty$ on $\tau$ (upper plot) and $f_j$ (lower plot), normalized by the power $|x_0(\infty)|^2$ in the no-jittering case.  For all other parameters, we use the Dark SRF values in \tabref{parameters}.  In blue and orange, we show numerical estimates of the expected power, for both Gaussian and DMP jittering, respectively, accompanied by shaded bands representing the total error $\sigma_{|x(t)|^2}$ on our estimate.  Each data point is estimated using $N=1000$ simulations, each with integration time $T=15\s$.  In black, we show the analytic result in \eqref{power}, which applies when $\alpha\ll1$.  The estimate applies well even outside this regime, and the Gaussian and DMP cases agree except in the case of significant power suppression when $f_j$ is very small.  (Note that the vertical axis in the lower plot utilizes a logarithmic scale.)  The dashed grey lines indicate the Dark SRF values for $\tau$ and $f_j$, respectively, shown in \tabref{parameters}.  These lie in the regime where the analytic estimate is reliable.}
    \label{fig:param_dependences}
\end{figure}
%%%%%%%%%%%%%%%%%%%%%%%%%%%%%%%%%%%%%%%%%%
%%%%%%%%%%%%%%%%%%%%%%%%%%%%%%%%%%%%%%%%%%

In \appref{shapiro}, we utilize the Shapiro-Loginov formuala to show that \eqref[s]{amplitude} and (\ref{eq:power}) also apply exactly in the ``pure DMP" case of DMP jittering with $\omega_j=0$.  If $\alpha\gg1$ the higher-point statistics of $\delta\omega(t)$ become important, so the results for Gaussian and DMP jittering differ.

The parameter $\rho$ defined in \eqref{rho} demonstrates how the timescale of jittering affects the suppression of power in the resonator.  Note that $\rho<1$, so long as $\gamma,\tau,\omega_j>0$.  When $\omega_j=0$ and $\tau\rightarrow\infty$, then the frequency of jittering is fixed to zero, and so we recover the fixed off-resonance case.  In this case, $\rho=1$ and \eqref{power} recovers the worst-case power suppression in \eqref{suppression} from a fixed frequency mismatch.  Smaller $\rho$ reduces the power suppression relative to the fixed off-resonance case.  In particular, in the $\tau\rightarrow0$ (broad-spectrum jittering) or $\omega_j\rightarrow\infty$ (high-frequency jittering) limit, $\rho\rightarrow0$ and the system accumulates power as if there were no jittering at all, even if $\delta\omega_0\gg\gamma$!  In general, we see that faster jittering reduces the power suppression.  In the next section, we will understand this phenomenon by considering the relative phase between the resonator and driving force.

In summary, we have found that power suppression may be negligible, even when $\delta\omega_0\gg\gamma$, so long as the jittering is sufficiently fast.  Generically, this condition is given by $\alpha\ll1$ [as defined in \eqref{alpha}], but we can make this condition more explicit in a few limits of interest.  In the case of monochromatic ($\tau\rightarrow\infty$) jittering, we find $\alpha=4\delta\omega^2/(\gamma^2+4\omega_j^2)$.  In other words, the power suppression is negligible, so long as the frequency of jittering is larger than the amplitude of jittering, $\omega_j\gg\delta\omega_0$.  In the case of broadband ($\omega_j=0$) jittering, we find $\alpha=4\delta\omega_0^2\tau/\gamma(2+\gamma\tau)$.  In this case, the power suppression is negligible, so long as $\tau\ll\gamma/2\delta\omega_0^2$.

In \figref{param_dependences}, we show the dependences of $\langle|x(t)|^2\rangle_\infty$ on $\tau$ and $\omega_j$.  We compare numerical results for both Gaussian and DMP jittering to the analytic formula in \eqref{power}.  As expected, this formula applies in the perturbative regime $\alpha\ll1$, as well as the pure DMP case, but it can even have good agreement outside these regimes.  We denote the Dark SRF parameter values by vertical dashed grey lines in this figure.  Numerically, we find that the power suppression (relative to $|x_0(\infty)|^2$) for Dark SRF is merely $13\%$.  The formula disagrees with the numerics in the case of small but nonzero $\omega_j$, where the power suppression becomes significant.  Note that the suppression in most of this regime is \emph{less} than the analytic result.  The Gaussian and DMP cases also begin to diverge in this regime, as higher-moment statistics of $\delta\omega(t)$ become more important.  While the dependence of the expected power on $\omega_j$ is monotonic (in the Gaussian and analytic cases), it exhibits a local minimum as a function of $\tau$ at
\begin{equation}
    \tau_\mathrm{min}=\frac2{\omega_j-\gamma},
\end{equation}
so long as $\omega_j>\gamma$.

%%%%%%%%%%%%%%%%%%%%%%%%%%%%%%%%%%%%%%%%%%%%%%%%%%%%%%%%%%%%%%%%%%%%%%%%%%%%%%%%%%%%%%%%%%
%%%%%%%%%%%%%%%%%%%%%%%%%%%%%%%%%%%%%%%%%%%%%%%%%%%%%%%%%%%%%%%%%%%%%%%%%%%%%%%%%%%%%%%%%%
\subsection{Power accumulation and relative phase}
%%%%%%%%%%%%%%%%%%%%%%%%%%%%%%%%%%%%%%%%%%%%%%%%%%%%%%%%%%%%%%%%%%%%%%%%%%%%%%%%%%%%%%%%%%
%%%%%%%%%%%%%%%%%%%%%%%%%%%%%%%%%%%%%%%%%%%%%%%%%%%%%%%%%%%%%%%%%%%%%%%%%%%%%%%%%%%%%%%%%%

%%%%%%%%%%%%%%%%%%%%%%%%%%%%%%%%%%%%%%%%%%
%%%%%%%%%%%%%%%%%%%%%%%%%%%%%%%%%%%%%%%%%%
\begin{figure}[t]
    \centering
    \includegraphics[width=0.95\columnwidth]{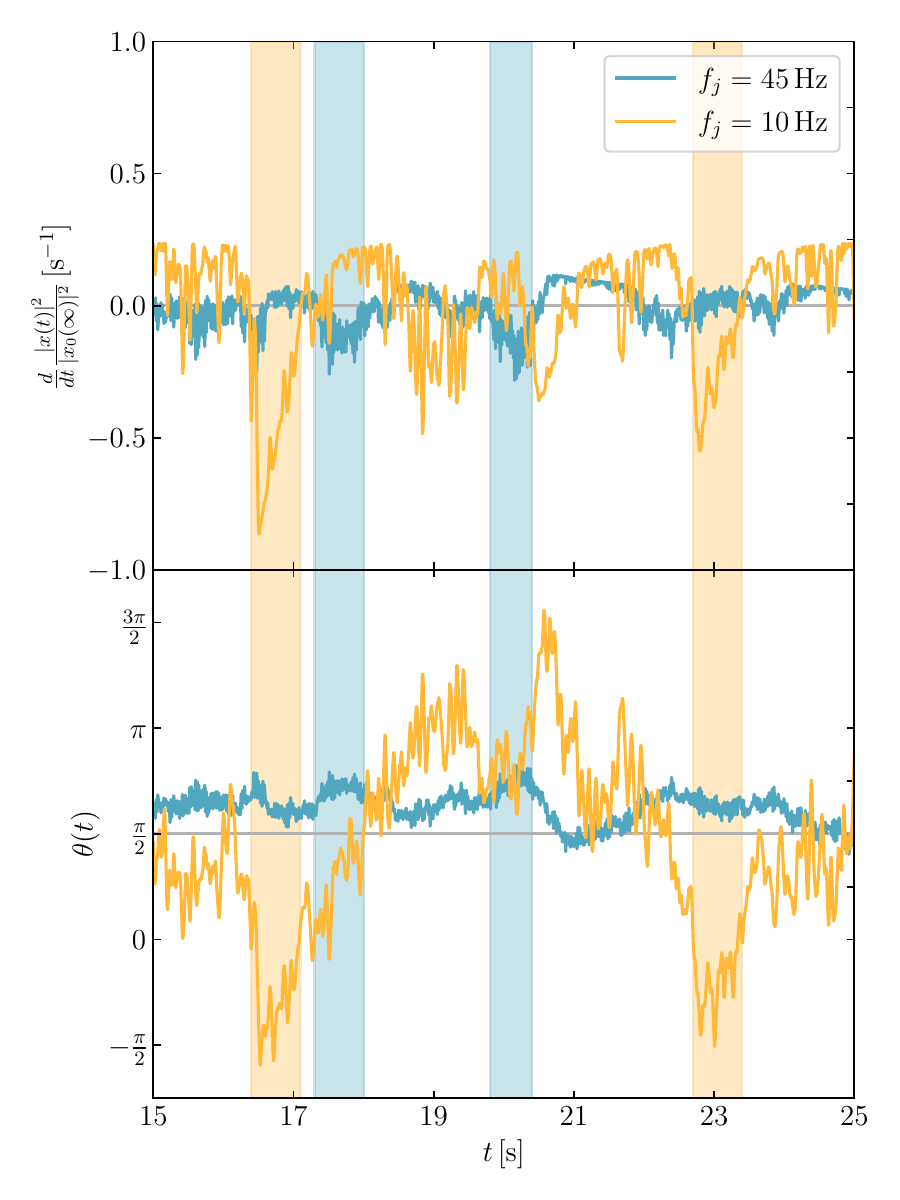}
    \caption{Relationship between power accumulation (upper plot) and relative phase $\theta$ (lower plot; see \eqref{phase} for definition), for two different values of $f_j$.  We use the values in \tabref{parameters} for all other parameters, and model the jittering as Gaussian.  We show the evolution of the system between $t=15\s$ and $t=25\s$.  We highlight a few periods of significant power loss using shaded bands.  (The band color matches the curve which is exhibiting power loss.)  Note that the periods of power loss occur when the relative phase deviates significantly from $\pi/2$.  The orange curve uses a smaller value of $f_j$ and so develops a larger relative phase.  This, in turn, leads to regions of more severe power loss.}
    \label{fig:phase}
\end{figure}
%%%%%%%%%%%%%%%%%%%%%%%%%%%%%%%%%%%%%%%%%%
%%%%%%%%%%%%%%%%%%%%%%%%%%%%%%%%%%%%%%%%%%

In the previous subsection, we saw that the rate of jittering affects how power accumulates in the resonator.  In this subsection, we will aim to understand this by examining the relationship between power accumulation and the relative phase between the resonator and driving force.  Let us begin by revisiting the evolution equations \eqref[s]{xevol} and (\ref{eq:xdotevol}).  In the limit where the time step is small, $\Delta t\ll\gamma,\delta\omega(t),\Delta\omega_F$, these can be rewritten as
\begin{align}
    \label{eq:yevol}
    y(t_{k+1})&=e^{-\frac{\gamma\Delta t}2+i(\omega_0+\delta\omega(t_k))\Delta t}y(t_k)-\frac{iF_0\Delta t}{\omega_0}e^{i\omega_F(t_k+\Delta t)},\\
    z(t_{k+1})&=e^{-\frac{\gamma\Delta t}2-i(\omega_0+\delta\omega(t_k))\Delta t}z(t_k),
    \label{eq:zevol}
\end{align}
where we have defined
\begin{align}
    y&=x-\frac{i\dot x}{\omega_0},\\
    z&=x+\frac{i\dot x}{\omega_0}.
\end{align}
It is clear from \eqref{zevol} that $|z(t)|$ will decay to zero after a time $t\gg1/\gamma$, irrespective of initial conditions.  Therefore at late times, $y\approx2x$.  Taking the absolute-value squared of \eqref{yevol} then implies that, at late times,
\begin{align}
    |x(t_{k+1})|^2=&(1-\gamma\Delta t)|x(t_k)|^2+\frac{\Delta t}{\omega_0}\IM[x(t_k)^*F_0e^{i\omega_Ft_k}]\nl
    +O(\Delta t^2).
    \label{eq:accum_discrete}
\end{align}
In the limit $\Delta t\rightarrow0$, this becomes
\begin{empheq}[box=\widefbox]{align}
    \frac d{dt}|x(t)|^2=-\gamma|x(t)|^2+\frac{|F_0||x(t)|}{\omega_0}\sin\theta(t),
    \label{eq:accumulation}
\end{empheq}
where $\theta(t)$ is the relative phase between the resonator and driving force
\begin{equation}
    \theta(t)=\arg\left[x(t)^*F_0e^{i\omega_Ft}\right].
    \label{eq:phase}
\end{equation}

\eqref{accumulation} gives us a clear understanding of how the power in a jittering resonator evolves.  The first term represents the power lost to dissipation, while the second term represents the power supplied by the driving force.  Importantly, this supplied power depends on the relative phase $\theta$.  In the no-jittering case with $\omega_F=\omega_0$, this phase is $\theta=\pi/2$ [see \eqref{nojittering_amp}], and the effect of the driving force is maximized.  In this case, the steady-state value is the one at which the power lost to dissipation equals the power supplied by the driving force. When $\omega_F\neq\omega_0$, then $\theta$ deviates from the optimal phase.  In particular, if $|\Delta\omega_F|\gg\gamma$ then $\theta$ approaches 0 or $\pi$ (depending on the sign of $\Delta\omega_F$) and the force ceases to be efficient in driving the resonator.  In summary, the reason that a fixed frequency mismatch leads to power suppression is because it causes the relative phase $\theta$ to deviate from $\pi/2$.

Jittering also causes a relative phase to develop.  From \eqref{yevol}, we see that a positive $\delta\omega(t)$ will decrease $\theta$, while a negative $\delta\omega(t)$ will increase it.  The important difference between jittering and a fixed frequency mismatch is that jittering changes sign, so that the accumulated phase can be washed out if the jittering is fast enough.  In particular, if $\tau$ is small, then $\delta\omega(t)$ randomizes on a short timescale, so that it spends very little time at a fixed value before potentially changing sign.  Likewise, if $\omega_j$ is large, then $\delta\omega(t)$ rapidly oscillates between positive and negative, leading to strong washout of the phase.  \eqref{accumulation} then implies that power will be accumulated efficiently, as in the no-jittering case.  This explains the dependence of $\rho$ on $\tau$ and $\omega_j$ and why a smaller $\rho$ leads to less power suppression.%
%%%%%%%%%%%%
\footnote{There is a related phenomenon in spectroscopy, known as Dicke narrowing~\cite{dicke1953}, where the linewidth of an emitting species becomes narrower when its mean free path is shorter than the wavelength of the light emitted.  In the language of this work, this can be stated as: when $\delta\omega_0\tau\ll1$ (and $\omega_j=0$), the bandwidth of the system is suppressed relative to the Doppler broadened width $2\delta\omega_0$.  This effect can also be understood in terms of the relative phase.  When $\omega_j=0$ and $\delta\omega_0\tau\ll1$, the phase performs a random walk, namely after time $T$, it has size $\sim\delta\omega_0\sqrt{\tau T}$.  In contrast, a frequency mismatch $\Delta\omega_F$ leads to a relative phase $\Delta\omega_FT$.  The former reaches $O(1)$ before the latter, so long as $\Delta\omega_F<\delta\omega_0^2\tau$.  In other words, the mismatch can be neglected for such frequencies, i.e., the system has bandwidth $2\delta\omega_0^2\tau$.  Note that when $\omega_j=0$ and $\gamma\tau\ll1$, \eqref{alpha} becomes $\alpha=2\delta\omega_0^2\tau/\gamma$, so Dicke narrowing recovers the rest-frame bandwidth $\gamma$ precisely in our perturbative limit!}
%%%%%%%%%%%%

In \figref{phase}, we demonstrate the relationship between power accumulation and the relative phase.  In particular, we show two cases with different values of $\omega_j$.  The lower $\omega_j$ implies a slower oscillation of the jittering, so that the system develops a larger relative phase.  This larger relative phase, in turn, leads to periods of more significant power loss.

%%%%%%%%%%%%%%%%%%%%%%%%%%%%%%%%%%%%%%%%%%%%%%%%%%%%%%%%%%%%%%%%%%%%%%%%%%%%%%%%%%%%%%%%%%
%%%%%%%%%%%%%%%%%%%%%%%%%%%%%%%%%%%%%%%%%%%%%%%%%%%%%%%%%%%%%%%%%%%%%%%%%%%%%%%%%%%%%%%%%%
\subsection{Power distribution and correlations}
\label{sec:distribution}
%%%%%%%%%%%%%%%%%%%%%%%%%%%%%%%%%%%%%%%%%%%%%%%%%%%%%%%%%%%%%%%%%%%%%%%%%%%%%%%%%%%%%%%%%%
%%%%%%%%%%%%%%%%%%%%%%%%%%%%%%%%%%%%%%%%%%%%%%%%%%%%%%%%%%%%%%%%%%%%%%%%%%%%%%%%%%%%%%%%%%

%%%%%%%%%%%%%%%%%%%%%%%%%%%%%%%%%%%%%%%%%%
%%%%%%%%%%%%%%%%%%%%%%%%%%%%%%%%%%%%%%%%%%
\begin{figure}[t]
    \centering
    \includegraphics[width=0.95\columnwidth]{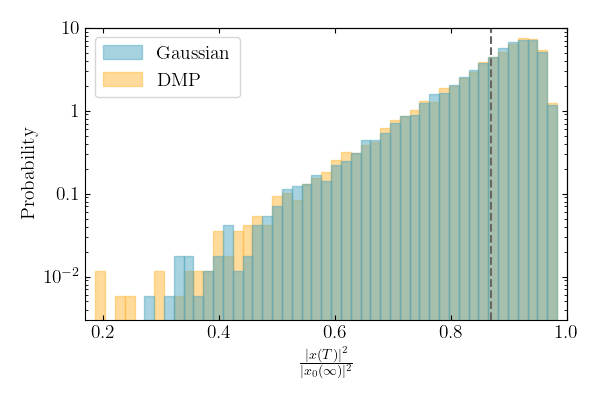}
    \caption{Probability distribution of $|x(T)|^2$ at $T=15\s$, normalized by the no-jittering power $|x_0(\infty)|^2$.  Here we use the parameter values in \tabref{parameters}.  In blue, we show the case of Gaussian jittering, while in orange we show the case of DMP jittering.  Each distribution consists of $N=10^4$ samples.  The dashed grey line indicates the mean power $\langle|x(t)|^2\rangle_\infty$.  Note that both cases exhibit similar distributions, which rise exponentially and then drop rapidly before the no-jittering value $|x_0(\infty)|^2$.}
    \label{fig:histogram}
\end{figure}
%%%%%%%%%%%%%%%%%%%%%%%%%%%%%%%%%%%%%%%%%%
%%%%%%%%%%%%%%%%%%%%%%%%%%%%%%%%%%%%%%%%%%

In \secref{power}, we discussed the expected power in the jittering resonator system.  As \figref{GPDMPcomparison} demonstrates, the real-time behavior of the power around this expected value can be quite complicated, with a nontrivial distribution and temporal correlations.  Here, we explore some of the other statistics that characterize the signal power.

First, we consider the distribution of $|x(t)|^2$ (at fixed time $t$) for different realizations of the jittering.  In \figref{histogram}, we show this distribution for both the cases of Gaussian and DMP jittering.  For the parameter values in \tabref{parameters}, these two cases exhibit very similar distributions, which rise exponentially until they peak near $\langle|x(t)|^2\rangle_\infty$, and then drop rapidly.  Note that the power in the system never exceeds the no-jittering case $|x_0(\infty)|^2$.

Next, we study the temporal correlations of the system.  Just as we did for the jittering in \eqref{jittering_corr}, we may define an autocorrelation function%
%%%%%%%%%%%%
\footnote{\label{ftnt:time_invariance}%
It is not immediately obvious that $C_x$ should only depend on the difference $t-t'$.  This would be the case, for instance, if \eqref{resonator} were time-translation invariant.  The asymptotic ensemble average $\langle\cdot\rangle_\infty$ ensures that the initial conditions and the $\delta\omega(t)$ term in this equation do not break time-translation invariance, but naively the driving force $F(t)$ does.  From \eqref{yevol}, one can show that, in the $t\rightarrow\infty$ limit, $\hat x(t)=x(t)e^{-i\omega_Ft}$ should satisfy the differential equation
\begin{equation}
    \dot{\hat x}=\left(-\frac\gamma2+i\left(\delta\omega(t)-\Delta\omega_F\right)\right)\hat x-\frac{iF_0}{2\omega_0}.
\end{equation}
The force term in this equation has no explicit time dependence and so $\langle\hat x(t)\hat x(t')\rangle$ should only depend on the difference $t-t'$.  It is then straightforward to see that $C_x$ should as well.}
%%%%%%%%%%%%
\begin{equation}
    C_x(t-t')\equiv\langle x(t)x(t')^*\rangle_\infty
    \label{eq:resonator_correlation}
\end{equation}
of the resonator.  From \eqref{amplitude}, we see that $x(t)$ and $x(t')$ individually have nonzero expectation values, and so a more instructive quantity will be the autocovariance function
\begin{equation}
    K_x(t-t')\equiv C_x(t-t')-\langle x(t)\rangle_\infty\langle x(t')^*\rangle_\infty,
    \label{eq:autocovariance}
\end{equation}
which goes to zero when $t$ and $t'$ are far separated.  Moreover, from \eqref{nojittering_amp}, it is not hard to see that the latter term in \eqref{autocovariance} will be proportional to $e^{i\omega_0(t-t')}$.  It is therefore helpful to define
\begin{equation}
    \hat K_x(t)=e^{-i\omega_0t}K_x(t)
\end{equation}
to remove this time dependence.  As we will see, $\hat K_x(t)$ is real and decays to zero at large $t$.

In \appref[ces]{perturbative} and \ref{app:shapiro}, we derive analytic expressions for $\hat K_x(t)$ of the form
\begin{equation}
    \hat K_x(t)=(c_1\cos(\omega_K|t|)+c_2\sin(\omega_K|t|))e^{-|t|/\tau_1}+c_3e^{-|t|/\tau_2},
    \label{eq:functional_form}
\end{equation}
in both the perturbative limit $\alpha\ll1$ [see \eqref{Kxhat_full}] and the pure DMP case [see \eqrefRange{beta1}{beta3}].  In the former limit, we find $\tau_1=\tau$, $\omega_K=\omega_j$, and $\tau_2=2/\gamma$.

\begin{figure}[t]
    \centering
    \includegraphics[width=0.95\columnwidth]{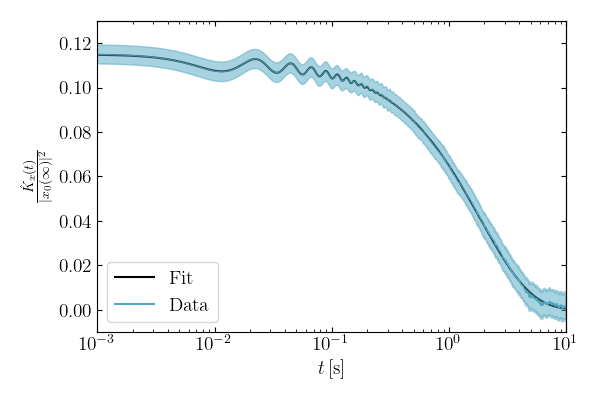}
    \caption{Correlation function $\hat K_x(t)$ for the parameter values in \tabref{parameters}. 
    In blue, we show our numerical estimate of $\hat K_x(t)$, along with a shaded band indicating the total error $\sigma_{\hat K_x(t)}$.  In black, we show a numerical fit of the form in \eqref{functional_form} [see \tabref{fit} for fit values].}
    \label{fig:autocorrelation}
\end{figure}

In \figref{autocorrelation}, we show $\hat K_x(t)$ for the parameter values in \tabref{parameters}.  To generate this numerical estimate, we evolve $N=10^4$ realizations of the system.  For each realization, we first evolve for $T=15\s$ to remove the transient contribution, and then compare $x(T)$ with $x(T+t)$ to estimate $\hat K_x(t)$.  We show the corresponding total error $\sigma_{\hat K_x(t)}$ as a shaded band around the mean estimate.  (Note that because all of these data points are derived from the same $N$ realizations, these errors are not independent.)  We also fit the data to
\eqref{functional_form}.  \tabref{fit} shows the timescales determined by the fit, in comparison to their values predicted by \eqref{Kxhat_full} and by \eqrefRange{beta1}{beta3}.  From \figref{autocorrelation}, we see that, for the parameter values in \tabref{parameters}, the final term in \eqref{functional_form} dominates, so that $\tau_2$ is the most relevant timescale on which correlations decay.  Generically though, any of $\tau_1$, $\omega_K$, or $\tau_2$ may be relevant for describing the temporal correlations of the system.

%%%%%%%%%%%%%%%%%%%%%%%%%%%%%%%%%%%%%%%%%%
%%%%%%%%%%%%%%%%%%%%%%%%%%%%%%%%%%%%%%%%%%
\begin{table}
    \centering
    \begin{tabular}{c c c c}
        \hline\hline
        Parameter&Perturbative&Pure DMP&Fit\\
        \hline
        $\tau_1$&$0.064\s$&$0.062\s$&$0.098\s$\\
        $f_K$&$45.0\Hz$&$45.1\Hz$&$44.7\Hz$\\
        $\tau_2$&$2.12\s$&$1.85\s$&$1.85\s$\\
        \hline\hline
    \end{tabular}
    \caption{Timescales appearing in the autocorrelation function $\hat K_x(t)$ [see \eqref{functional_form}].  In the first column, we list the values for the timescales predicted by \eqref{Kxhat_full}, which applies in the perturbative limit $\alpha\ll1$.  In the second column, we show the corresponding values from \eqrefRange{beta1}{beta3}, which apply in the pure DMP case (DMP jittering with $\omega_j=0$).  In the final column, we show the numerical fit corresponding to the black line in \figref{autocorrelation}.  In all cases, we use the parameter values in \tabref{parameters}.}
    \label{tab:fit}
\end{table}
%%%%%%%%%%%%%%%%%%%%%%%%%%%%%%%%%%%%%%%%%%
%%%%%%%%%%%%%%%%%%%%%%%%%%%%%%%%%%%%%%%%%%

%%%%%%%%%%%%%%%%%%%%%%%%%%%%%%%%%%%%%%%%%%%%%%%%%%%%%%%%%%%%%%%%%%%%%%%%%%%%%%%%%%%%%%%%%%
%%%%%%%%%%%%%%%%%%%%%%%%%%%%%%%%%%%%%%%%%%%%%%%%%%%%%%%%%%%%%%%%%%%%%%%%%%%%%%%%%%%%%%%%%%
\section{Spectral features and sensitivity}
\label{sec:spectral}
%%%%%%%%%%%%%%%%%%%%%%%%%%%%%%%%%%%%%%%%%%%%%%%%%%%%%%%%%%%%%%%%%%%%%%%%%%%%%%%%%%%%%%%%%%
%%%%%%%%%%%%%%%%%%%%%%%%%%%%%%%%%%%%%%%%%%%%%%%%%%%%%%%%%%%%%%%%%%%%%%%%%%%%%%%%%%%%%%%%%%

Now that we have discussed many of the characteristics of the jittering resonator system, we wish to determine its sensitivity to detect a signal.  In many sensing contexts, this is characterized by a ``signal-to-noise ratio" (SNR) which compares the response of the system to a monochromatic driving force vs. its response to broadband noise.  A strongly resonant system can achieve a large SNR for even very small signals by leveraging the coherence properties of the signal as compared to the noise.  That is, if a resonator is driven by a nearly monochromatic signal, its response will remain correlated for a very long time, whereas if it is driven by white noise, its response will exhibit no temporal correlations.  Many common formulae for the SNR therefore rely not only on the expected signal power, but also on the ``coherence time" of the signal.

As we have seen in \secref{distribution}, the temporal correlations of a jittering resonator can depend on multiple timescales, and so it is difficult to define a single ``coherence time" for the system response.  Rather than considering the time-domain properties of the system, in order to define a SNR, it will be more appropriate to analyze the system in the frequency domain.  In this section, we will characterize the spectral features of the system and use them to compute a SNR.  In particular, we will derive the spectrum of the system in response to a monochromatic driving force and compare this to the spectrum in response to white noise.  We will find that jittering can introduce sidebands into both spectra.  Moreover, we will show that, in the perturbative limit, the sensitivity of a jittering resonator is the same as a stable resonator.

%%%%%%%%%%%%%%%%%%%%%%%%%%%%%%%%%%%%%%%%%%
%%%%%%%%%%%%%%%%%%%%%%%%%%%%%%%%%%%%%%%%%%
%%%%%%%%%%%%%%%%%%%%%%%%%%%%%%%%%%%%%%%%%%
%%%%%%%%%%%%%%%%%%%%%%%%%%%%%%%%%%%%%%%%%%

%%%%%%%%%%%%%%%%%%%%%%%%%%%%%%%%%%%%%%%%%%%%%%%%%%%%%%%%%%%%%%%%%%%%%%%%%%%%%%%%%%%%%%%%%%
%%%%%%%%%%%%%%%%%%%%%%%%%%%%%%%%%%%%%%%%%%%%%%%%%%%%%%%%%%%%%%%%%%%%%%%%%%%%%%%%%%%%%%%%%%
\subsection{Response spectrum}
\label{sec:response}
%%%%%%%%%%%%%%%%%%%%%%%%%%%%%%%%%%%%%%%%%%%%%%%%%%%%%%%%%%%%%%%%%%%%%%%%%%%%%%%%%%%%%%%%%%
%%%%%%%%%%%%%%%%%%%%%%%%%%%%%%%%%%%%%%%%%%%%%%%%%%%%%%%%%%%%%%%%%%%%%%%%%%%%%%%%%%%%%%%%%%

Let us return to the system of a jittering resonator, described by \eqref{resonator}, driven by a monochromatic force, described by \eqref{force}.  However, instead of analyzing the system as a function of time, let us consider its properties as a function of frequency.  That is, we will be interested in the Fourier transform
\begin{equation}
    \tilde x(f)=\int dt\,x(t)e^{-2\pi ift}
\end{equation}
of the system response.  If the force is monochromatic with frequency $\omega_f=2\pi f_F$, its Fourier transform is simply
\begin{equation}
    \tilde F(f)=\int dt\,F(t)e^{-2\pi ift}=F_0\delta(f-f_F).
\end{equation}
In the absence of jittering, a monochromatic force leads to a monochromatic response, that is
\begin{equation}
    \tilde x_0(f)=\frac{F_0\delta(f-f_F)}{\omega_0(i\gamma-2\Delta\omega_F)}=\chi_0(f)\tilde F(f).
    \label{eq:response_nojit}
\end{equation}
Here we have introduced the mechanical susceptibility (in the absence of jittering)
\begin{equation}
    \chi_0(\omega)=\frac1{\omega_0(i\gamma-2\Delta\omega)},
\end{equation}
where $\Delta\omega=\omega-\omega_0=2\pi(f-f_0)$.  By linearity, \eqref{response_nojit} also applies for non-monochromatic forces with Fourier transform $\tilde F(f)$.

We will be interested in the response of the system not only to a signal, but also to noise. 
 The latter is an ensemble of forces, which can be characterized by its force PSD $S_F$.  This is defined as
\begin{equation}
    S_F(f_F)\delta(f_F-f_F')=\langle\tilde F(f_F)\tilde F(f'_F)^*\rangle_F.
    \label{eq:forcePSD}
\end{equation}
Note that $\langle\cdot\rangle_F$ here denotes an ensemble average over realizations of the force, rather than over realizations of the jittering (which we denote by $\langle\cdot\rangle_\infty$).  In this work, we will primarily consider noise sources whose force PSDs are flat as a function of frequency
\begin{equation}
    S_F^\mathrm{noise}(f_F)=S_F^N,
\end{equation}
and will refer to such sources as \emph{thermal} noise.  A monochromatic signal may also be described by a force PSD
\begin{equation}
    S_F^\mathrm{sig}(f_F)=|F_0|^2\delta(f_F-f_{F,0}).
    \label{eq:monochromaticPSD}
\end{equation}
We will also be interested in signals which are not exactly monochromatic, but have some linewidth $\gamma_F$.  These are described by a force PSD
\begin{equation}
    S_F^\mathrm{sig}(\omega_F)=\frac{4|F_0|^2\gamma_F}{\gamma_F^2+4(\omega_F-\omega_{F,0})^2}
\end{equation}
(which approaches \eqref{monochromaticPSD} in the limit $\gamma_F\rightarrow0$.)

\begin{figure}[p]
    \centering
    \includegraphics[width=0.95\columnwidth]{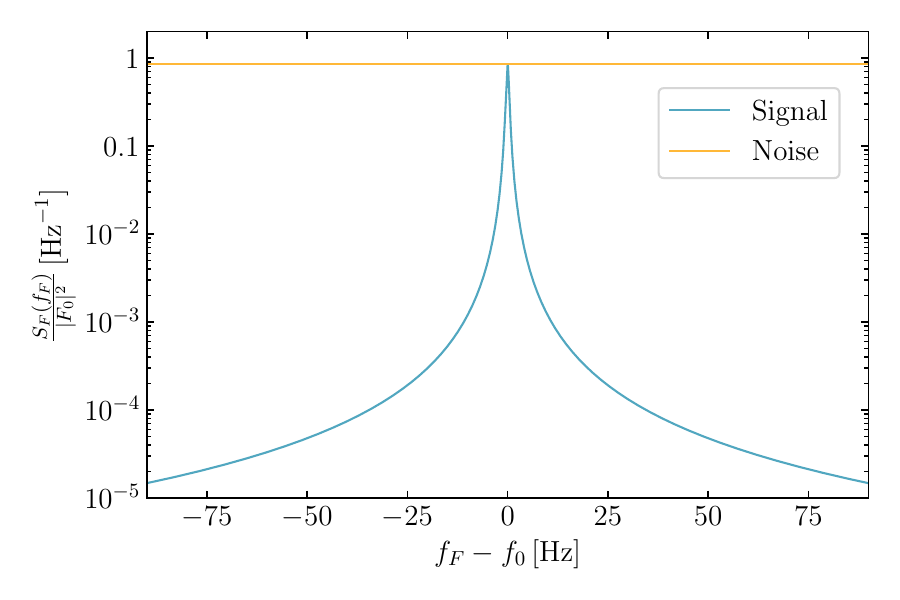}
    \includegraphics[width=0.95\columnwidth]{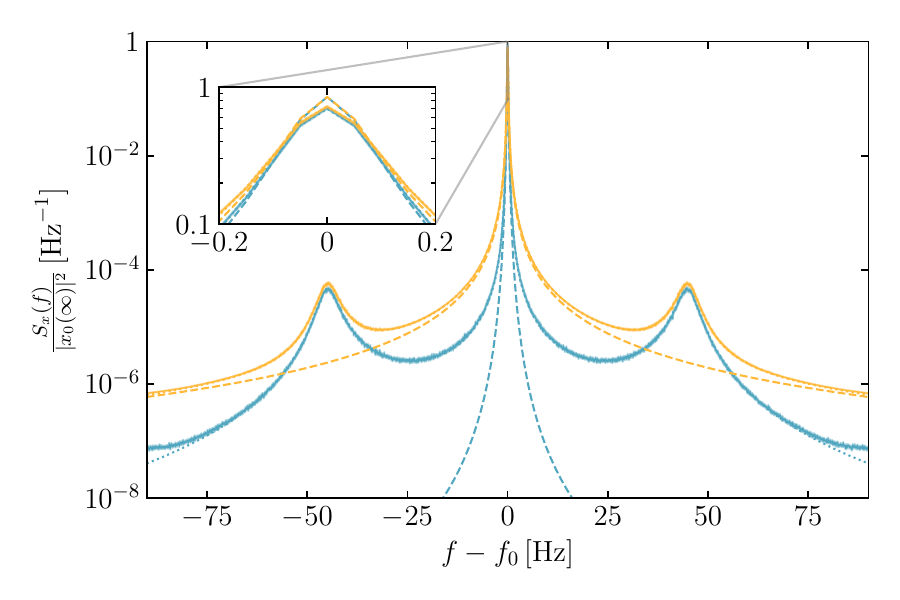}
    \includegraphics[width=0.9\columnwidth]{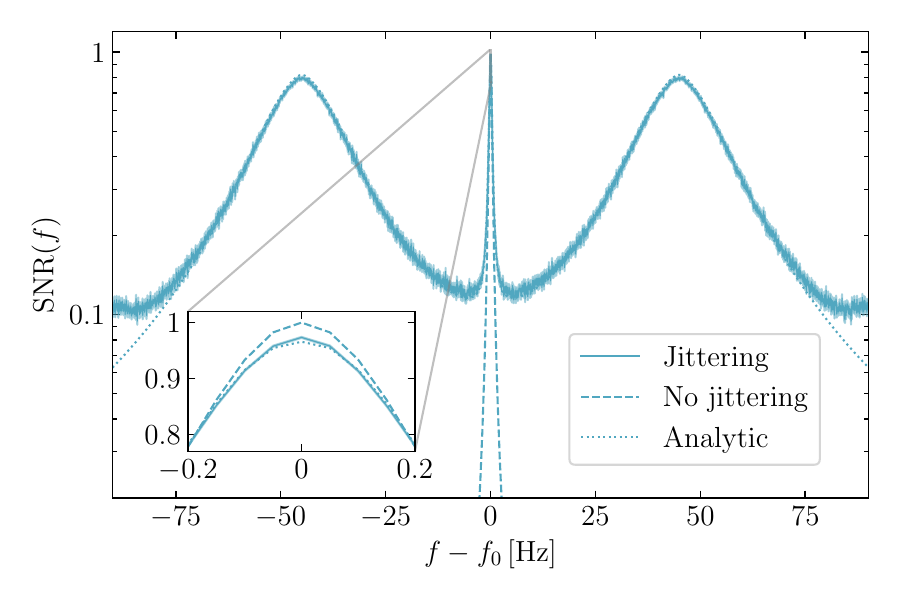}
    \caption{Force, position, and SNR spectra.  In the upper plot, we show the force PSD $S_F(f)$ for a narrowband signal (blue) with linewidth $\gamma_F=2\pi\cdot0.75\Hz$ and for broadband noise (orange).  In the middle plot, we show the resulting position PSD $S_x(f)$ for the parameter values in \tabref{parameters}.  In the lower plot, we show the SNR as a function of frequency.  In the lower two plots, numerical estimates are shown as a solid line with a shaded error band; the no-jittering case is shown as a dashed line; and the perturbative result derived using \eqref{mechanicalPSD} is shown as a dotted line.  We also show inset plots of the behavior near $f_0$ to demonstrate the slight degradation caused by jittering.}
    \label{fig:snr}
\end{figure}

Similarly to \eqref{forcePSD}, we may define a position PSD for the response of the system
\begin{equation}
    S_x(f)\delta(f-f')=\langle\tilde x(f)\tilde x(f')^*\rangle_F.
    \label{eq:monochromatic_positionPSD}
\end{equation}
From \eqref{response_nojit}, we see that, in the absence of jittering, a force PSD $S_F$ will result in a corresponding position PSD
\begin{equation}
    S_x(f)=|\chi_0(f)|^2S_F(f).
\end{equation}
Note that even if the force PSD of a thermal noise source is flat, the position PSD will be peaked near $f_0$ (see \figref{snr}).

Now we wish to understand how jittering will affect the position PSD of the system.  We saw that in the absence of jittering, a monochromatic force will lead to a monochromatic response.  When we include jittering, this will not necessarily be the case.  For a monochromatic force $\tilde F(f)=F_0\delta(f-f_F)$ and fixed realization of jittering, let the response of the system be given by%
%%%%%%%%%%%%
\footnote{Note that our definition of $\chi$ is such that $\chi(f_F,f)=\chi_0(f)\delta(f-f_F)$ in the no-jittering case.}
%%%%%%%%%%%%
\begin{equation}
    \tilde x(f)=\chi(f_F,f)F_0.
    \label{eq:monochromatic_chi}
\end{equation}
The mechanical susceptibility $\chi(f_F,f)$ here depends on the particular realization of the jittering.  We may consider asymptotic ensemble-averaged statistics of $\chi(f_F,F)$, as we did for other quantities in \secref{characteristics}.  Since the statistics of $x(t)$ are time-translation invariant (see footnote~\ref{ftnt:time_invariance}), then the Fourier modes $\tilde x(f)$ and $\tilde x(f')$ are statistically independent.  This will be similarly true for $\chi(f_F,f)$, and so we may define a PSD for the mechanical susceptibility
\begin{equation}
    S_\chi(f_F,f)\delta(f-f')=\langle\chi(f_F,f)\chi(f_F,f')^*\rangle_\infty.
    \label{eq:Schi}
\end{equation}

In the case of a generic force $\tilde F(f_F)$, \eqref{monochromatic_chi} can be generalized to
\begin{equation}
    \tilde x(f)=\int df_F\,\chi(f_F,f)\tilde F(f_F).
\end{equation}
If we have a force ensemble described by the PSD $S_F(f_F)$, this can be converted into a position PSD by averaging over both the jittering and over the force PSD (denoted by $\langle\cdot\rangle_{F,\infty}$)
\begin{widetext}
\begin{align}
    \label{eq:positionPSD1}
    \langle\tilde x(f)\tilde x(f')^*\rangle_{F,\infty}&=\int df_Fdf'_F\,\langle\chi(f_F,f)\chi(f'_F,f')^*\rangle_\infty\langle F(f_F)F(f'_F)^*\rangle_F\\\label{eq:positionPSD2}
    &=\int df_F\,\langle\chi(f_F,f)\chi(f_F,f')^*\rangle_\infty S_F(f_F)\\
    &=\delta(f-f')\int df_F\,S_\chi(f_F,f)S_F(f_F).
    \label{eq:positionPSD3}
\end{align}
\end{widetext}
Note that averaging over the jittering is essential, in order to make the Fourier modes statistically independent.  Unlike the no-jittering case in \eqref{monochromatic_positionPSD}, a position PSD cannot be defined by averaging only over realizations of the force.  In summary, \eqref{positionPSD3} implies that the position PSD for the system can be computed as
\begin{equation}
    S_x(f)=\int df_F\,S_\chi(f_F,f)S_F(f_F).
    \label{eq:convolve}
\end{equation}
The essential physics of the jittering resonator is captured by the mechanical susceptibility PSD $S_\chi(f_F,f)$.  Once this has been calculated, we can convert any force PSD into a corresponding position PSD.

In \appref{perturbative}, we show that in the perturbative limit $\alpha\ll1$, the mechanical susceptibility PSD is given by%
%%%%%%%%%%%%
\footnote{In \eqref{Schi_analytic}, we omit subleading corrections to the coefficient of the delta function.  See \eqref{mechanicalPSD} for full first-order expression.  We use this full expression for the analytic curves in the lower two plots of \figref{snr}.}
%%%%%%%%%%%%
\begin{empheq}[box=\widefbox]{align}
    S_\chi(\omega_F,\omega)=\frac{\delta(f-f_F)+\frac{4S_{\delta\omega}(f-f_F)}{\gamma^2+4\Delta\omega^2}}{\omega_0^2(\gamma^2+4\Delta\omega_F^2)}.
    \label{eq:Schi_analytic}
\end{empheq}
Note that this expression does not rely on the form of the jittering PSD $S_{\delta\omega}$, and applies even if it is not described by a Lorentzian [as in \eqref{Sdelomega}].  \eqref{Schi_analytic} indicates that there are two contributions to the response of the system to a monochromatic driving force: a response at the driving frequency $f=f_F$, which is typically present in a stable resonator; and a broadband jittering-induced response, which occurs at all frequencies.  Importantly, the denominator in \eqref{Schi_analytic} implies that, when the system is driven on-resonance $\Delta\omega_F<\gamma$, \emph{both} contributions are enhanced.  This implies that a narrowband signal can drive the system more efficiently than broadband noise over a wide range of response frequencies.

In \figref{snr}, we numerically estimate $S_\chi(f_F,f)$ and use it to compute the response PSD $S_x(f)$.  The upper plot of \figref{snr} shows two driving force PSDs: a narrowband signal with linewidth $\gamma_F=2\pi\cdot0.75\Hz$ comparable to the emitter cavity in the Dark SRF experiment~\cite{Romanenko2023}; and a broadband noise source, which is normalized to match the peak of the signal force PSD, $S_F^N=4|F_0|^2/\gamma_F$.  To compute $S_\chi(f_F,f)$, for each $f_F$, we simulate \eqref{resonator} with a monochromatic driving force $N=100$ times.  The first $T=15\s$ of each simulation are discarded to remove transient features, and the next $t_\mathrm{int}=20\s$ are Fourier transformed to find $\chi(f_F,f)$ for that realization of the jittering.  The resulting susceptibilities are then averaged as in \eqref{Schi} to compute $S_\chi(f_F,f)$.  In the middle plot of \figref{snr}, we use this $S_\chi(f_F,f)$ to compute $S_x(f)$ for both the signal and noise force PSDs via \eqref{convolve}.  We show the total errors $\sigma_{S_x(f)}$ on our estimates as shaded bands.  In the lower plot, we show the ratio between the signal and noise response PSDs.

%%%%%%%%%%%%%%%%%%%%%%%%%%%%%%%%%%%%%%%%%%
%%%%%%%%%%%%%%%%%%%%%%%%%%%%%%%%%%%%%%%%%%
%%%%%%%%%%%%%%%%%%%%%%%%%%%%%%%%%%%%%%%%%%
%%%%%%%%%%%%%%%%%%%%%%%%%%%%%%%%%%%%%%%%%%

We see from \figref{snr} that jittering introduces richer spectral features into the response of the system.  In particular, the Lorentzian jittering we have assumed in this work introduces two sidebands at $f=f_0\pm f_j$ to both the signal and noise responses.  In the time domain, these sidebands correspond to the oscillatory contributions to $\hat K_x(t)$ we saw in \eqref{functional_form}.  Notably, the SNR in these sidebands is comparable to the SNR near $f_0$.  We note that jittering slightly degrades both the signal and noise spectra at $f=f_0$.  This is expected since we have seen that the total power in the system (which in the frequency domain, is dominated by the response on-resonance) is slightly suppressed by jittering.  Finally, we note that the perturbative analytic prediction for $S_x(f)$ agrees well with our numerical estimate, except near $f=f_0\pm2f_j$.  Just as jittering produces features at $f_0\pm f_j$ at first order, we observe that it produces features at $f_0\pm2f_j$ at second order, which are not captured by our perturbative calculations.

%%%%%%%%%%%%%%%%%%%%%%%%%%%%%%%%%%%%%%%%%%%%%%%%%%%%%%%%%%%%%%%%%%%%%%%%%%%%%%%%%%%%%%%%%%
%%%%%%%%%%%%%%%%%%%%%%%%%%%%%%%%%%%%%%%%%%%%%%%%%%%%%%%%%%%%%%%%%%%%%%%%%%%%%%%%%%%%%%%%%%
\subsection{Signal-to-noise ratio}
\label{sec:SNR}
%%%%%%%%%%%%%%%%%%%%%%%%%%%%%%%%%%%%%%%%%%%%%%%%%%%%%%%%%%%%%%%%%%%%%%%%%%%%%%%%%%%%%%%%%%
%%%%%%%%%%%%%%%%%%%%%%%%%%%%%%%%%%%%%%%%%%%%%%%%%%%%%%%%%%%%%%%%%%%%%%%%%%%%%%%%%%%%%%%%%%

Now that we have understood how to compute the spectrum of the system response, we wish to characterize the sensitivity of the system.  Let us first understand how to define an appropriate SNR in the frequency domain.  Suppose that we have performed an experiment of duration $t_\mathrm{int}$ (which does not include any initial evolution required to remove transient contributions).  We can then compute the Fourier transform $\tilde x(f_i)$ of the response at discrete frequencies $f_i=i/t_\mathrm{int}$.  Each of these values represents an independent measurement.  Let us suppose that our experiment has integrated for long enough that we can resolve all features in both the signal and noise response spectra, that is, $2\pi/t_\mathrm{int}\ll\gamma,\gamma_F,\omega_j,2/\tau$.  Then the SNR for a single measurement $\tilde x(f_i)$ is simply
\begin{equation}
    \mathrm{SNR}(f_i)=\frac{S_x^\mathrm{sig}(f_i)}{S_x^\mathrm{noise}(f_i)}.
\end{equation}
In \appref{SNR}, we show that, when the response $\tilde x(f)$ is Gaussian, the total SNR can be computed by summing the SNR from these independent measurements in quadrature
\begin{equation}
    \mathrm{SNR}_\mathrm{tot}^2=\sum_i\mathrm{SNR}(f_i)^2=t_\mathrm{int}\int df\left(\frac{S_x^\mathrm{sig}(f)}{S_x^\mathrm{noise}(f)}\right)^2.
    \label{eq:totalSNR}
\end{equation}

In the no-jittering case, we see from \figref{snr} that the SNR is highest near $f_0$.  When jittering is present, wide sidebands appear which also exhibit large SNR.  One may expect from \eqref{totalSNR} that a jittering resonator is more sensitive, as these sidebands could dominate the total SNR.  However, as we show in \appref{perturbative}, the response in the sidebands is highly non-Gaussian, and therefore it is not appropriate to apply \eqref{totalSNR}.  The responses at different frequencies in these sidebands are highly correlated, as a single large driving force $\tilde F(f_F)$ at $f_F\approx f_0$ can lead to an increase in $\tilde x(f)$ across several frequencies within the sideband.  This naturally leads to highly non-Gaussian behavior.  The response near $f_0$, however, remains Gaussian.  Therefore, \eqref{totalSNR} can be applied in the frequency range around the central peak.  In the perturbative limit $\alpha\ll1$, this will yield a sensitivity comparable to the no-jittering case.  Specifically, in both cases, we find
\begin{equation}
    \mathrm{SNR}_\mathrm{tot}\sim\sqrt{t_\mathrm{int}\cdot\frac{\gamma_F}{2\pi}\cdot\left(\frac{4|F_0|^2}{\gamma_FS_F^N}\right)^2}.
\end{equation}

%%%%%%%%%%%%%%%%%%%%%%%%%%%%%%%%%%%%%%%%%%%%%%%%%%%%%%%%%%%%%%%%%%%%%%%%%%%%%%%%%%%%%%%%%%
%%%%%%%%%%%%%%%%%%%%%%%%%%%%%%%%%%%%%%%%%%%%%%%%%%%%%%%%%%%%%%%%%%%%%%%%%%%%%%%%%%%%%%%%%%
\section{Discussion and Dark SRF bound}
\label{sec:discussion}
%%%%%%%%%%%%%%%%%%%%%%%%%%%%%%%%%%%%%%%%%%%%%%%%%%%%%%%%%%%%%%%%%%%%%%%%%%%%%%%%%%%%%%%%%%
%%%%%%%%%%%%%%%%%%%%%%%%%%%%%%%%%%%%%%%%%%%%%%%%%%%%%%%%%%%%%%%%%%%%%%%%%%%%%%%%%%%%%%%%%%

%%%%%%%%%%%%%%%%%%%%%%%%%%%%%%%%%%%%%%%%%%
%%%%%%%%%%%%%%%%%%%%%%%%%%%%%%%%%%%%%%%%%%
\begin{figure}[t]
    \centering
    \includegraphics[width=0.95\columnwidth]{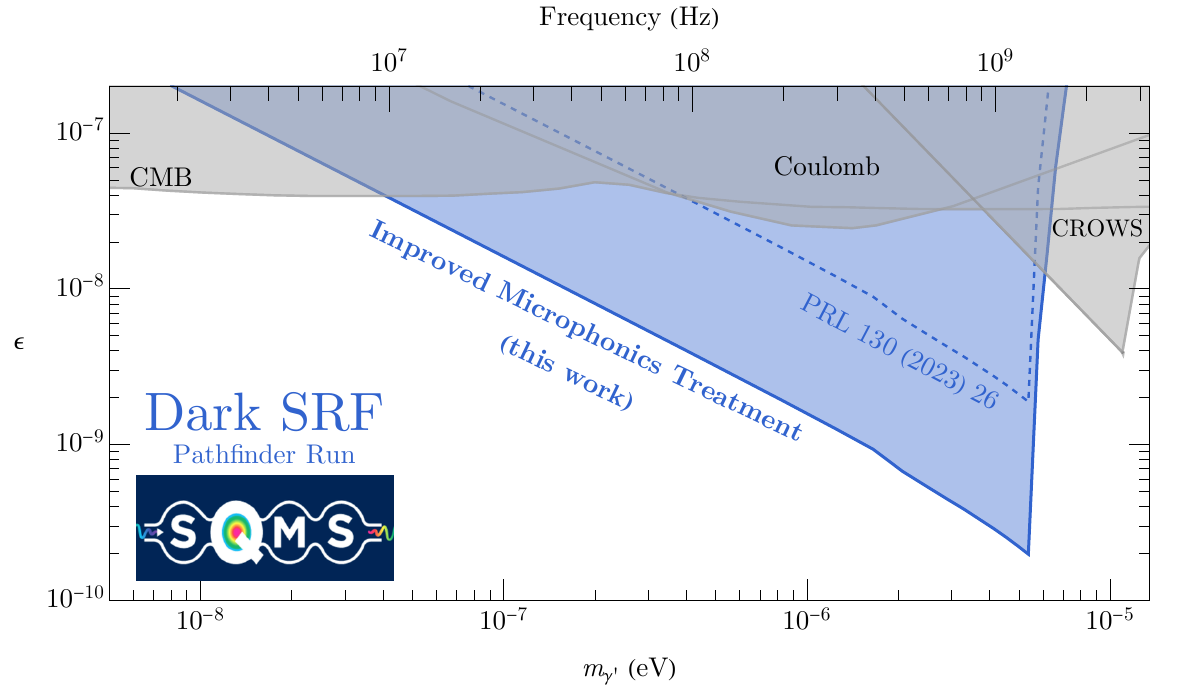}
    \caption{Refined dark-photon exclusion bound, based on the Dark SRF pathfinder run, derived in an accompanying letter~\cite{darksrf} which utilizes our improved jittering treatment. The new result (solid line) improves on the originally reported result (dashed line)~\cite{Romanenko2023} by about an order of magnitude.}
    \label{fig:DarkSRF}
\end{figure}
%%%%%%%%%%%%%%%%%%%%%%%%%%%%%%%%%%%%%%%%%%
%%%%%%%%%%%%%%%%%%%%%%%%%%%%%%%%%%%%%%%%%%

In this work, we analyzed the effects of jittering in resonant systems.  While, naively, one may expect jittering to suppress the signal power in a resonant system, we have shown that a jittering resonator can accumulate power as efficiently as a stable resonator.  In particular, we highlighted the importance of the jittering timescale.  We showed that if this timescale is short, then no relative phase develops between the resonator and driving force, and power accumulates as if there were no jittering.  In particular, in the case of monochromatic jittering, the frequency of jittering must exceed its amplitude ($\omega_j\gg\delta\omega_0$), while in the case of broadband jittering, the correlation time $\tau$ must be less than $\gamma/2\delta\omega_0^2$.  We also demonstrated that jittering introduces distinct spectral features into the response of the system, while preserving the central resonant peak.  This implies that a jittering resonator can be as sensitive as a stable resonator.

Our results have important consequences for existing and future resonant experiments with high quality factors, such as the Dark SRF search.  In \citeR{Romanenko2023}, a dark-photon exclusion bound was calculated for Dark SRF, assuming a power suppression factor of $7.7\times10^{-6}$ due to jittering.  In \secref{characteristics}, we showed that the true suppression is only $0.87$!  The existing Dark SRF data, therefore, translates to a much stronger constraint than previously reported.  In an accompanying letter~\cite{darksrf}, we utilize the results of this work to improve the exclusion bound on the dark-photon kinetic mixing parameter $\epsilon$ originally reported in \citeR{Romanenko2023}.

The refined dark-photon exclusion bound from the Dark SRF experiment, which properly accounts for jittering, is shown in \figref{DarkSRF}.  This updated bound improves on the originally reported bound by over an order of magnitude.  As the sensitivity to $\epsilon$ scales as $\sim\mathrm{SNR}^{1/4}$, this corresponds to an enhancement of the SNR by four orders of magnitude!  This improved Dark SRF pathfinder-run result is the world-leading constraint on non-dark-matter dark photons over a wide range of masses below $6\,\rm \mu eV$.  This also translates to the best laboratory-based limits on the photon mass $m_\gamma < 1.6\times 10^{-15}\rm eV = 2.9\times 10^{-48}\,\rm g$!  The improved jittering treatment derived in this work enables the full utilization of high-$Q$ devices, like Dark SRF, for new-physics searches.

%%%%%%%%%%%%%%%%%%%%%%%%%%%%%%%%%%%%%%%%%%%%%%%%%%%%%%%%%%%%%%%%%%%%%%%%%%%%%%%%%%%%%%%%%%
%%%%%%%%%%%%%%%%%%%%%%%%%%%%%%%%%%%%%%%%%%%%%%%%%%%%%%%%%%%%%%%%%%%%%%%%%%%%%%%%%%%%%%%%%%
\acknowledgments

We thank Peter Graham, Roni Harnik, and Harikrishnan Ramani for helpful discussions regarding the application of our results to future experiments.  We also thank Dmitry Budker and Kevin Zhou for bringing our attention to discussions of similar effects in the literature.   We thank Joshua Foster for highlighting the importance of non-Gaussianities in the system response to us.  We acknowledge Stephen Henrich for his contribution to this work in its initial stage.

S.K. and Z.L. are supported in part by the U.S. Department of Energy (DOE), Office of Science, National Quantum Information Science Research Centers, Superconducting Quantum Materials and Systems Center (SQMS) under contract number DE-AC02-07CH11359, and in part by the DOE grant DE-SC0011842 and a Sloan Research Fellowship from the Alfred P. Sloan Foundation at the University of Minnesota.

Some of the computing for this project was performed on the Sherlock cluster. We would like to thank Stanford University and the Stanford Research Computing Center for providing computational resources and support that contributed to these research results.

The code used for this research is made publicly available through Github~\cite{github} under CC-BY-NC-SA.

%%%%%%%%%%%%%%%%%%%%%%%%%%%%%%%%%%%%%%%%%%%%%%%%%%%%%%%%%%%%%%%%%%%%%%%%%%%%%%%%%%%%%%%%%%
%%%%%%%%%%%%%%%%%%%%%%%%%%%%%%%%%%%%%%%%%%%%%%%%%%%%%%%%%%%%%%%%%%%%%%%%%%%%%%%%%%%%%%%%%%

%%%%%%%%%%%%%%%%%%%%%%%%%%%%%%%%%%%%%%%%%%%%%%%%%%%%%%%%%%%%%%%%%%%%%%%%%%%%%%%%%%%%%%%%%%
%%%%%%%%%%%%%%%%%%%%%%%%%%%%%%%%%%%%%%%%%%%%%%%%%%%%%%%%%%%%%%%%%%%%%%%%%%%%%%%%%%%%%%%%%%
\bibliographystyle{JHEP}
\bibliography{references.bib}
%%%%%%%%%%%%%%%%%%%%%%%%%%%%%%%%%%%%%%%%%%%%%%%%%%%%%%%%%%%%%%%%%%%%%%%%%%%%%%%%%%%%%%%%%%
%%%%%%%%%%%%%%%%%%%%%%%%%%%%%%%%%%%%%%%%%%%%%%%%%%%%%%%%%%%%%%%%%%%%%%%%%%%%%%%%%%%%%%%%%%

%%%%%%%%%%%%%%%%%%%%%%%%%%%%%%%%%%%%%%%%%%%%%%%%%%%%%%%%%%%%%%%%%%%%%%%%%%%%%%%%%%%%%%%%%%
%%%%%%%%%%%%%%%%%%%%%%%%%%%%%%%%%%%%%%%%%%%%%%%%%%%%%%%%%%%%%%%%%%%%%%%%%%%%%%%%%%%%%%%%%%
\appendix
\onecolumngrid
%%%%%%%%%%%%%%%%%%%%%%%%%%%%%%%%%%%%%%%%%%%%%%%%%%%%%%%%%%%%%%%%%%%%%%%%%%%%%%%%%%%%%%%%%%
%%%%%%%%%%%%%%%%%%%%%%%%%%%%%%%%%%%%%%%%%%%%%%%%%%%%%%%%%%%%%%%%%%%%%%%%%%%%%%%%%%%%%%%%%%

%%%%%%%%%%%%%%%%%%%%%%%%%%%%%%%%%%%%%%%%%%%%%%%%%%%%%%%%%%%%%%%%%%%%%%%%%%%%%%%%%%%%%%%%%%
%%%%%%%%%%%%%%%%%%%%%%%%%%%%%%%%%%%%%%%%%%%%%%%%%%%%%%%%%%%%%%%%%%%%%%%%%%%%%%%%%%%%%%%%%%
\section{Perturbative results}
\label{app:perturbative}
%%%%%%%%%%%%%%%%%%%%%%%%%%%%%%%%%%%%%%%%%%%%%%%%%%%%%%%%%%%%%%%%%%%%%%%%%%%%%%%%%%%%%%%%%%
%%%%%%%%%%%%%%%%%%%%%%%%%%%%%%%%%%%%%%%%%%%%%%%%%%%%%%%%%%%%%%%%%%%%%%%%%%%%%%%%%%%%%%%%%%

In the following two appendices, we derive analytic results for various asymptotic ensemble-averaged quantities.  In this appendix, we adopt a perturbative approach to solve \eqref{resonator} [with a monochromatic force, as in \eqref{force}].  That is, we will assume that $x(t)\approx x_0(t)$ to leading order at all times $t$.  Note that this is not an ensemble-averaged assumption [i.e., $\langle x(t)\rangle_\infty\approx x_0(t)$], but rather we assume this for each realization of jittering.  We will derive the condition for this approximation to hold, and then compute various ensemble-averaged quantities in this approximation.  In this appendix, we will not assume $\omega_F=\omega_0$, so the results here will be more general than those quoted in \secref[s]{power} and \ref{sec:distribution}.

Let us expand $x(t)\approx x_0(t)+x_1(t)+x_2(t)$ to second order.  The zeroth order contribution to \eqref{resonator} is simply
\begin{equation}
    \ddot x_0(t)+\gamma\dot x_0(t)+\omega_0^2x_0(t)=F_0e^{i\omega_Ft},
\end{equation}
which is solved by the asymptotic expression in \eqref{nojittering_amp}.  It will be helpful for us to compute solutions in the frequency domain.  Let us define
\begin{align}
    \tilde x_n(f)&=\int_0^T dt\,x_n(t)e^{-2\pi ift}\\
    \widetilde{\delta\omega}(f)&=\int_0^T dt\,\delta\omega(t)e^{-2\pi ift},
\end{align}
where $T$ is the total integration time.  In the frequency domain, \eqref{nojittering_amp} becomes%
%%%%%%%%%%%%
\footnote{\eqref[s]{x0sol}, (\ref{eq:x1sol}) and (\ref{eq:x2sol}) form a specific solution to \eqref{resonator}, which does not necessarily satisfy the inital boundary conditions we have fixed.  The full solution should consist of the these contributions $x_n^{(0)}$, along with a homogeneous contribution $x_n^{(h)}$ which satisfies \eqref{resonator} with no driving force, $F(t)=0$.  The homogenous contribution should be chosen so that the full solution $x_n=x_n^{(0)}+x_n^{(h)}$ satisfies the initial conditions.  Generically, we expect $x_n^{(h)}(t)$ to decay for $t\gg2/\gamma$.  Therefore, in the limit $T\rightarrow\infty$, its Fourier tranform $\tilde x_n^{(h)}(f)$ vanishes, and so $\tilde x_n(f)\rightarrow\tilde x_n^{(0)}(f)$.}
%%%%%%%%%%%%
\begin{equation}
    \lim_{T\rightarrow\infty}\tilde x_0(f)=\frac{F_0\delta(f-f_F)}{\omega_0(i\gamma-2\Delta\omega_F)}.
    \label{eq:x0sol}
\end{equation}

The first-order contribution to \eqref{resonator} is
\begin{equation}\label{eq:first-order}
    \ddot x_1(t)+\gamma\dot x_1(t)+\omega_0^2x_1(t)=-2\omega_0\delta\omega(t)x_0(t).
\end{equation}
The left-hand side here arises from plugging $x\rightarrow x_1$ into \eqref{resonator}, while the right-hand side is the first-order contribution which arises from plugging $x\rightarrow x_0$  into \eqref{resonator}.  The solution to \eqref{first-order} in the frequency domain is
\begin{equation}
    \lim_{T\rightarrow\infty}\tilde x_1(f)=-\int df'\,\frac{2\widetilde{\delta\omega}(f')\tilde x_0(f-f')}{i\gamma-2\Delta\omega}=-\frac{2F_0\widetilde{\delta\omega}(f-f_F)}{\omega_0(i\gamma-2\Delta\omega_F)(i\gamma-2\Delta\omega)},
    \label{eq:x1sol}
\end{equation}
where $\Delta\omega=2\pi(f-f_0)$.  Note that because $\langle\widetilde{\delta\omega}(f)\rangle=0$, then $\langle\tilde x_1(f)\rangle_\infty=0$.  Therefore, if we are interested in quantities like $\langle x(t)\rangle_\infty$, we must compute the solution to second order.  At this order, \eqref{resonator} becomes
\begin{equation}
    \ddot x_2(t)+\gamma\dot x_2(t)+\omega_0^2x_2(t)=-2\omega_0\delta\omega(t)x_1(t).
\end{equation}
(Note that we still assume $\delta\omega(t)\ll\omega_0$, so that terms such as $\delta\omega(t)^2x_0(t)$ which might contribute to the RHS of this equation are negligible.)  This is solved by
\begin{equation}
    \lim_{T\rightarrow\infty}\tilde x_2(f)=\frac{4F_0}{\omega_0(i\gamma-2\Delta\omega_F)(i\gamma-2\Delta\omega)}\int df'\,\frac{\widetilde{\delta\omega}(\Delta f-f')\widetilde{\delta\omega}(f'-\Delta f_F)}{i\gamma-2\omega'},
    \label{eq:x2sol}
\end{equation}
where $\omega'=2\pi f'$.

Equipped with \eqref[s]{x1sol} and (\ref{eq:x2sol}), we may now compute ensemble-averaged quantities.  First, let us determine when our perturbative approximation actually holds.  This will occur when $x_1(t)\ll x_0(t)$ for all $t$.  Again, we require this to hold not only in expectation, but for all realizations.  We should therefore demand that $\langle |x_1(t)|^2\rangle_\infty\ll|x_0(\infty)|^2$.  From \eqref{x1sol}, we can compute
\begin{align}
    \langle|x_1(t)|^2\rangle_\infty&=\int dfdf'\,\langle\tilde x_1(f)\tilde x_1(f')^*\rangle_\infty e^{2\pi i(f-f')t}\\
    &=\frac{4|F_0|^2}{\omega_0^2(\gamma^2+4\Delta\omega_F^2)}\int df\,\frac{S_{\delta\omega}(f)}{\gamma^2+4(\omega+\Delta\omega_F)^2}=\frac{4\delta\omega_0^2}{\gamma^2}\rho_1\cdot|x_0(\infty)|^2,
    \label{eq:amplitude_perturbative}
\end{align}
where
\begin{equation}
    \rho_1=\frac{\gamma\tau(2+\gamma\tau)\left[(2+\gamma\tau)^2+4\omega_j^2\tau^2+4\Delta\omega_F^2\tau^2\right]}{\left[(2+\gamma\tau)^2+4(\omega_j-\Delta\omega_F)^2\tau^2\right]\left[(2+\gamma\tau)^2+4(\omega_j+\Delta\omega_F)^2\tau^2\right]}.
\end{equation}
We see that the perturbative approximation holds when $\alpha\equiv4\delta\omega^2/\gamma^2\cdot\rho_1\ll1$.

Now we can compute the corrections to $\langle x(t)\rangle_\infty$.  As mentioned above, the leading correction comes at second order
\begin{equation}
    \langle x_2(t)\rangle_\infty=\frac{4F_0e^{i\omega_Ft}}{\omega_0(i\gamma-2\Delta\omega_F)^2}\int df\,\frac{S_{\delta\omega}(f)}{i\gamma-2\omega-2\Delta\omega_F}=-\frac{4\delta\omega_0^2}{\gamma^2}\rho_2\cdot\lim_{t\rightarrow\infty}x_0(t),
    \label{eq:x2_perturbative}
\end{equation}
where
\begin{equation}
    \rho_2=\frac\gamma{\gamma+2i\Delta\omega_F}\cdot\frac{\gamma\tau(2+\gamma\tau+2i\Delta\omega_F\tau)}{(2+\gamma\tau+2i\Delta\omega_F\tau)^2+4\omega_j^2\tau^2}.
\end{equation}
The expected power can then easily be computed as
\begin{align}
    \langle|x(t)|^2\rangle_\infty&=|x_0(\infty)|^2+\langle|x_1(t)|^2\rangle_\infty
    +2\RE\left[\lim_{t\rightarrow\infty}x_0(t)^*\cdot\langle x_2(t)\rangle_\infty\right]\\
    &=\left(1+\frac{4\delta\omega_0^2}{\gamma^2}(\rho_1-2\RE[\rho_2])\right)|x_0(\infty)|^2.
    \label{eq:power_perturbative}
\end{align}
Note that $\rho_1$ is real, although $\rho_2$ is not necessarily.  When $\omega_F=\omega_0$, then $\rho_1=\rho_2=\rho$ as defined in \eqref{rho}.  In this case, \eqref[s]{amplitude} and (\ref{eq:power}) readily follow from \eqref[s]{amplitude_perturbative} and (\ref{eq:power_perturbative}) [in the limit $\alpha\ll1$].  \eqref[s]{amplitude_perturbative} and (\ref{eq:x2_perturbative}) can also be applied to more general jittering spectra.  Notably, these expressions are linear in $S_{\delta\omega}$, so that if a spectrum exhibits multiple peaks, $\alpha$ can be computed additively.

Now let us compute the autocovariance function $\hat K_x(t)$.  In the perturbative limit, this is given by
\begin{align}
    \hat K_x(s)&=e^{-i\omega_0s}\left(\langle x(t+s)x(t)^*\rangle_\infty-\langle x(t+s)\rangle_\infty\langle x(t)^*\rangle_\infty\right)=e^{-i\omega_0s}\langle x_1(t+s)x_1(t)^*\rangle_\infty\\
    &=\frac{4|F_0|^2}{\omega_0^2(\gamma^2+4\Delta\omega_F^2)}\int df\,\frac{S_{\delta\omega}(f)e^{2\pi ifs}}{\gamma^2+4(\omega+\Delta\omega_F)^2}.
\end{align}
The full expression for $\hat K_x(t)$ is quite complex.  Here we simply quote the result in the $\omega_F=\omega_0$ case
\begin{equation}
    \hat K_x(t)=|x_0(\infty)|^2\cdot\frac{4\delta\omega_0^2}{\gamma^2}\cdot\frac{\gamma^2\tau^2\left[(\gamma^2\tau^2-4+4\omega_j^2\tau^2)\cos(\omega_j|t|)+8\omega_j\tau\sin(\omega_j|t|)\right]e^{-|t|/\tau}+2\gamma\tau[4-\gamma^2\tau^2+4\omega_j^2\tau^2]e^{-\gamma|t|/2}}{[(2-\gamma\tau)^2+4\omega_j^2\tau^2][(2+\gamma\tau)^2+4\omega_j^2\tau^2]}.
    \label{eq:Kxhat_full}
\end{equation}
Note that, in this case, $\hat K_x(t)$ is real and $\hat K_x(0)=\langle|x_1(t)|^2\rangle_\infty$.  Moreover, \eqref{Kxhat_full} has the form of \eqref{functional_form} with $\tau=\tau_1$, $\omega_K=\omega_j$, and $\tau_2=2/\gamma$.

We also derive the mechanical susceptibility PSD.  From its definition in \eqref{Schi}, it is given by
\begin{equation}
    S_\chi(f_F,f)=\frac{\langle\tilde x(f)\tilde x(f')^*\rangle_\infty}{|F_0|^2\delta(f-f')}=\frac{\left(1-\frac{8\delta\omega^2}{\gamma^2}\RE[\rho_2]\right)\delta(f-f_F)+\frac{4S_{\delta\omega}(f-f_F)}{\gamma^2+4\Delta\omega^2}}{\omega_0^2(\gamma^2+4\Delta\omega_F^2)}.
    \label{eq:mechanicalPSD}
\end{equation}
The delta function term here comes from the $|\tilde x_0(f)|^2$ and $2\RE[\tilde x_0(f)^*\langle\tilde x_2(f)\rangle_\infty]$ contributions to this expression, while the latter term comes from the $\langle|\tilde x_1(f)|^2\rangle_\infty$ contribution.  In the time domain, the latter contribution corresponds to $\hat K_x(t)$, while the former corresponds to the constant term which was subtracted off from $C_x(t)$ in the definition of $\hat K_x(t)$ [see \eqref{autocovariance}].

Finally, let us comment on the Gaussianity of the response $\tilde x(f)$.  For a generic force $\tilde F(f_F)$, to first order, the response can be written as
\begin{equation}
    \tilde x(f)=\int df_F\,\chi(f_F,f)\tilde F(f_F)=\int df_F\,\frac{\tilde F(f_F)}{\omega_0(i\gamma-2\Delta\omega_F)}\left[\delta(f-f_F)-\frac{2\widetilde{\delta\omega}(f-f_F)}{i\gamma-2\Delta\omega}\right].
    \label{eq:response_nlo}
\end{equation}
From \figref{snr}, we can see that near the resonant peak $f\approx f_0$, the response matches that of the no-jittering case.  In other words, at these frequencies, the former term in \eqref{response_nlo} dominates.  In this case, $\tilde x(f)$ is proportional to $\tilde F(f)$, so that as long as the latter is assumed to be Gaussian, the former will be as well.  This means that we may apply Wick's theorem, i.e. for $f_1,f_2,f_3,f_4\approx f_0$,
\begin{equation}
    \langle\tilde x(f_1)\tilde x(f_2)^*\tilde x(f_3)\tilde x(f_4)^*\rangle_{F,\infty}\approx\langle\tilde x(f_1)\tilde x(f_2)^*\rangle_{F,\infty}\langle\tilde x(f_3)\tilde x(f_4)^*\rangle_{F,\infty}+\langle\tilde x(f_1)\tilde x(f_4)^*\rangle_{F,\infty}\langle\tilde x(f_3)\tilde x(f_2)^*\rangle_{F,\infty}.
\end{equation}
On the other hand, for frequencies in the sidebands of \figref{snr}, i.e. $f\approx f_0\pm f_j$, the response is dominated by the contribution from jittering.  This response arises from on-resonant forces $f_F\approx f_0$ which are up/down-converted by jittering.  This implies that the latter term in \eqref{response_nlo} dominates in this frequency range.  In this case, $\tilde x(f)$ is a product of Gaussian variables (if the jittering is assumed to be Gaussian), and so will not obey Gaussian statistics.  More specifically, for $f_1,f_2,f_3,f_4\approx f_0\pm f_j$, it can be shown that
\begin{align}
    \langle\tilde x(f_1)\tilde x(f_2)^*\tilde x(f_3)\tilde x(f_4)^*\rangle_{F,\infty}\approx2\big[&\langle\tilde x(f_1)\tilde x(f_2)^*\rangle_{F,\infty}\langle\tilde x(f_3)\tilde x(f_4)^*\rangle_{F,\infty}+\langle\tilde x(f_1)\tilde x(f_4)^*\rangle_{F,\infty}\langle\tilde x(f_3)\tilde x(f_2)^*\rangle_{F,\infty}\nl
    +\langle\tilde x(f_1)\tilde x(2f_0-f_3)^*\rangle_{F,\infty}\langle\tilde x(f_2)\tilde x(2f_0-f_4)^*\rangle_{F,\infty}\big].
    \label{eq:nonGaussian}
\end{align}

%%%%%%%%%%%%%%%%%%%%%%%%%%%%%%%%%%%%%%%%%%%%%%%%%%%%%%%%%%%%%%%%%%%%%%%%%%%%%%%%%%%%%%%%%%
%%%%%%%%%%%%%%%%%%%%%%%%%%%%%%%%%%%%%%%%%%%%%%%%%%%%%%%%%%%%%%%%%%%%%%%%%%%%%%%%%%%%%%%%%%
\section{Shapiro-Loginov approach}
\label{app:shapiro}
%%%%%%%%%%%%%%%%%%%%%%%%%%%%%%%%%%%%%%%%%%%%%%%%%%%%%%%%%%%%%%%%%%%%%%%%%%%%%%%%%%%%%%%%%%
%%%%%%%%%%%%%%%%%%%%%%%%%%%%%%%%%%%%%%%%%%%%%%%%%%%%%%%%%%%%%%%%%%%%%%%%%%%%%%%%%%%%%%%%%%

In this appendix, we derive analytic results by utilizing the Shapiro-Loginov formula~\cite{shapiro1978formulae}, which computes derivatives of certain ensemble-averaged quantities of Gaussian processes or DMPs.  This approach will allow us to construct coupled differential equations for various ensemble-averaged quantities.  Generically, this sequence of coupled equations will be infinite, but in the DMP case, we will find that the sequence can be truncated.  Moreover, in order to derive closed form results, we will have to assume that some of the ensemble-averaged quantities are small, so their differential equations can be neglected.  There are a couple notable cases where this assumption is justified, namely, the perturbative case and the DMP case with $\omega_j=0$.  (In the perturbative case, the distinction between Gaussian and DMP jittering is negligible, so that the earlier DMP assumption is not critical.)  The results in this appendix are, therefore, directly applicable in these two cases.  However, the numerical simulations in \secref{power} suggest that they may even hold in a wider range of scenarios.  Throughout this appendix, we take $\omega_F=\omega_0$ and we suppress the $\infty$ subscript in all ensemble averages $\langle\cdot\rangle_\infty$.

%%%%%%%%%%%%%%%%%%%%%%%%%%%%%%%%%%%%%%%%%%%%%%%%%%%%%%%%%%%%%%%%%%%%%%%%%%%%%%%%%%%%%%%%%%
%%%%%%%%%%%%%%%%%%%%%%%%%%%%%%%%%%%%%%%%%%%%%%%%%%%%%%%%%%%%%%%%%%%%%%%%%%%%%%%%%%%%%%%%%%
\subsection{Amplitude and power}
\label{app:shapiro_amppower}
%%%%%%%%%%%%%%%%%%%%%%%%%%%%%%%%%%%%%%%%%%%%%%%%%%%%%%%%%%%%%%%%%%%%%%%%%%%%%%%%%%%%%%%%%%
%%%%%%%%%%%%%%%%%%%%%%%%%%%%%%%%%%%%%%%%%%%%%%%%%%%%%%%%%%%%%%%%%%%%%%%%%%%%%%%%%%%%%%%%%%

We begin by computing the expected resonator amplitude $\langle x(t)\rangle$ and power $\langle|x(t)|^2\rangle$.  The Shapiro-Loginov formula~\cite{shapiro1978formulae} states that given a Gaussian process or DMP $\eta(t)$ with autocorrelation function as in \eqref{etaphi2}, any functional $y[\eta(t)]$ satisfies
\begin{equation}
    \left(\frac{d}{dt}+\frac{1}{\tau}\right)^n\langle \eta(t)y(t)\rangle=\left\langle \eta(t)\frac{d^n}{dt^n}y(t)\right\rangle,
    \label{eq:Shapiro Loginov}
\end{equation}
for any integer $n$.  One can verify, for example, that \eqref{Shapiro Loginov} is fulfilled when $y(t)$ is a polynomial functional of $\eta(t)$.  For two independent processes $\eta(t)$ and $\phi(t)$ with the same correlation time $\tau$, we have the generalized Shapiro-Loginov formula~\cite{burov2016noisy}, 
\begin{equation}
    \left(\frac{d}{dt}+\frac{2}{\tau}\right)^n\langle \eta(t)\phi(t)y(t)\rangle=\left\langle \eta(t)\phi(t)\frac{d^n}{dt^n}y(t)\right\rangle.
    \label{eq:Shapiro Loginov 2}
\end{equation}

In order to apply the Shapiro-Loginov formula, we replace $\delta\omega(t)$ in \eqref{resonator} by the decomposition in \eqref{decomposition} to find
\begin{equation}
    \ddot x(t)+\gamma\dot x(t) + \omega_0^2 x(t) +2\omega_0[ \eta (t)\cos (2\pi f_{j} t)+\phi (t)\sin (2\pi f_{j} t)] x(t) = F_0 e^{i \omega_0 t},
    \label{eq:decomp_resonator}
\end{equation}
in the limit of large $\omega_0$.  Taking the ensemble average of \eqref{decomp_resonator} will give us a differential equation for $\langle x(t)\rangle$, in terms of $\langle\eta(t)x(t)\rangle$ and $\langle\phi(t)x(t)\rangle$.  In order to arrive at differential equations for these quantities, we can first multiply \eqref{decomp_resonator} by $\eta(t)$ or $\phi(t)$, take an ensemble average, and then apply \eqref{Shapiro Loginov}.  These differential equations will in turn involve new ensemble-averaged quantities, for which we will also require differential equations.  In order to truncate this infinite sequence of equations, we restrict ourselves to the case where $\eta(t)$ and $\phi(t)$ are DMPs.  By definition, these processes satisfy
\begin{equation}
    \langle\eta(t)\eta(t)y(t)\rangle=\langle\phi(t)\phi(t)y(t)\rangle=\delta\omega_0^2\langle y(t)\rangle
    \label{eq:DMP_property}
\end{equation}
for any functional $y[\eta(t),\phi(t)]$.  In this case, the above procedure will yield a closed system of four coupled differential equations
\begin{align}
    \label{eq:x_diffeq}
    & L_0 \langle x(t)\rangle +2\omega_0\langle\eta(t)x(t)\rangle\cos(\omega_j t)+2\omega_0\langle\phi(t)x(t)\rangle\sin(\omega_j t)=F_0e^{i\omega_0 t},\\
    & L_1 \langle \eta(t)x(t)\rangle +2\omega_0\delta\omega_0^2\langle x(t)\rangle\cos(\omega_j t)+2\omega_0 \langle \eta(t)\phi(t)x(t)\rangle \sin (\omega_j t)=0,\label{eq:etax_diffeq}\\
    & L_1 \langle \phi(t)x(t)\rangle+2\omega_0 \langle \eta(t)\phi(t)x(t)\rangle \cos (\omega_j t)+2\omega_0\delta\omega_0^2\langle x(t)\rangle\sin(\omega_j t)=0,\label{eq:phix_diffeq}\\
    & L_2 \langle \eta(t)\phi(t)x(t)\rangle +2\omega_0\delta\omega_0^2\langle \phi(t)x(t)\rangle \cos (\omega_j t) +2\omega_0\delta\omega_0^2\langle \eta(t)x(t)\rangle\sin(\omega_j t)=0,
    \label{eq:etaphix_diffeq}
\end{align}
where we have defined the operators
\begin{equation}
    L_j\equiv\left(\frac{d}{dt}+\frac{j}{\tau}\right)^2+\gamma\left(\frac{d}{dt}+\frac{j}{\tau}\right)+\omega_0^2.
\end{equation}
Note that the RHS of \eqrefRange{etax_diffeq}{etaphix_diffeq} vanish because $\langle\eta(t)\rangle=\langle\phi(t)\rangle=\langle\eta(t)\phi(t)\rangle=0$.

The ensemble-averaged amplitude $\langle x(t)\rangle$ can be computed by solving \eqrefRange{x_diffeq}{etaphix_diffeq}.  In general, we are not aware of a closed form solution to this set of equations.  Instead, in this appendix, we make the simplifying assumption that $\langle\eta(t)\phi(t)x(t)\rangle\ll\delta\omega_0^2\langle x(t)\rangle$.  This allows us to neglect this quantity in \eqref[s]{etax_diffeq} and (\ref{eq:phix_diffeq}) and decouple \eqref{etaphix_diffeq}, reducing the system to
\begin{align}
    \label{eq:x_diffeq2}
    & L_0 \langle x(t)\rangle +2\omega_0\langle\eta(t)x(t)\rangle\cos(\omega_j t)+2\omega_0\langle\phi(t)x(t)\rangle\sin(\omega_j t)=F_0e^{i\omega_0 t},\\
    & L_1 \langle \eta(t)x(t)\rangle +2\omega_0\delta\omega_0^2\langle x(t)\rangle\cos(\omega_j t)=0,\label{eq:etax_diffeq2}\\
    & L_1 \langle \phi(t)x(t)\rangle+2\omega_0\delta\omega_0^2\langle x(t)\rangle\sin(\omega_j t)=0.
    \label{eq:phix_diffeq2}
\end{align}

There are two cases of note where this assumption is applicable.  One is the pure DMP case where $\omega_j=0$, as in this case $\delta\omega(t)=\eta(t)$, and so $\phi(t)$ decouples entirely.  The other is the perturbative case addressed in \appref{perturbative}. \eqref{x0sol} indicates that $x_0$ is independent of $\delta\omega(t)$, while \eqref{x1sol} indicates that $x_1$ is linear in $\delta\omega(t)$.  Because $\eta(t)$ and $\phi(t)$ both have mean zero, this implies
\begin{equation}
    \langle\eta(t)x_0(t)\rangle=\langle\phi(t)x_0(t)\rangle=\langle\eta(t)\phi(t)x_1(t)\rangle=0.
\end{equation}
In other words, $\langle x(t)\rangle$ is a zeroth-order quantity in the perturbative expansion, $\langle\eta(t)x(t)\rangle$ and $\langle\phi(t)x(t)\rangle$ are first-order, and $\langle\eta(t)\phi(t)x(t)\rangle$ is second-order.  This justifies the assumption to neglect the latter.  Note that the same logic also implies
\begin{equation}
    \langle\eta(t)\eta(t)x(t)\rangle=\langle\eta(t)\eta(t)x_0(t)\rangle+\langle\eta(t)\eta(t)x_2(t)\rangle=\delta\omega_0^2\langle x(t)\rangle+\mathcal O(\alpha^2)
\end{equation}
[and likewise for $\langle\phi(t)\phi(t)x(t)\rangle$], so that \eqref{DMP_property} can be applied in the perturbative case, even if the jittering is not modeled by a DMP.

Now let us solve \eqrefRange{x_diffeq2}{phix_diffeq2} for $\langle x(t)\rangle$.  It will be useful to note that, to leading order in $\omega_0$,
\begin{align}
    L_1[\langle \eta(t)x(t)\rangle \cos(\omega_j t)]&\approx L_1[\langle \eta(t)x(t)\rangle]\cos(\omega_j t)+2\frac{d}{dt}\langle \eta(t)x(t)\rangle\frac{d}{dt}\cos(\omega_j t)\\
    &=L_1[\langle \eta(t)x(t)\rangle]\cos(\omega_j t)-2i\omega_0\omega_j\langle \eta(t)x(t)\rangle\sin(\omega_j t)
    \label{eq:L1cos}
\end{align}
(and likewise for similar quantities).  Then by applying the operator $L_1L_1$ to \eqref{x_diffeq2}, we find
\begin{align}
    L_1L_1[F_0e^{i\omega_0t}]&=L_1L_1L_0 \langle x(t)\rangle +2\omega_0L_1L_1\left[\langle\eta(t)x(t)\rangle\cos(\omega_j t)+\langle\phi(t)x(t)\rangle\sin(\omega_j t)\right]\\
    &=(L_1 L_1 L_0 -4\omega_0^2\delta\omega_0^2L_1 )\langle x(t)\rangle+4i\omega_0^2\omega_jL_1[-\langle \eta(t)x(t)\rangle\sin(\omega_j t) +\langle \phi(t)x(t)\rangle\cos(\omega_j t)]\\
   &=(L_1 L_1 L_0 -4\omega_0^2\delta\omega_0^2L_1 )\langle x(t)\rangle+8\omega_0^3\omega_j^2 \left[\langle \eta(t)x(t)\rangle\cos(\omega_j t) +\langle \phi(t)x(t)\rangle\sin(\omega_j t)\right]\\
   &=(L_1 L_1 L_0 -4\omega_0^2\delta\omega_0^2L_1 )\langle x(t)\rangle+4\omega_0^2\omega_j^2 \left(F_0e^{i\omega_0 t}-L_0\langle x(t)\rangle\right)
   \label{eq:system_solution}
\end{align}
Since we are interested in the asymptotic solution of this system, we may apply the ansatz $\langle x(t)\rangle\propto e^{i\omega_0t}$.  In this case, we can replace $L_0\rightarrow i\gamma\omega_0$ and $L_1\rightarrow i\omega_0(\gamma+\frac2\tau)$ in \eqref{system_solution} [to leading order in $\omega_0$].  This yields the solution
\begin{align}
    \langle x(t)\rangle&=\frac{(L_1L_1-4\omega_0^2\omega_j^2)F_0e^{i\omega_0t}}{L_1L_1L_0-4\omega_0^2\delta\omega_0^2L_1-4\omega_0^2\omega_j^2L_0}=\frac{F_0e^{i\omega_0t}}{L_0-4\omega_0^2\delta\omega_0^2\cdot\frac{L_1}{L_1L_1-4\omega_0^2\omega_j^2}}\\
    &=\frac{F_0e^{i\omega_0t}}{i\gamma\omega_0\left(1+\frac{4\delta\omega_0^2}{\gamma^2}\cdot\frac{\gamma\tau(2+\gamma\tau)}{(2+\gamma\tau)^2+4\omega_j^2\tau^2}\right)}\equiv\frac{F_0e^{i\omega_0t}}{i\gamma\omega_0(1+\alpha)}.
    \label{eq:shapiro_amplitude}
\end{align}
The solutions to \eqref{etax_diffeq2} and (\ref{eq:phix_diffeq2}) are then
\begin{align}
    \label{eq:etax_average}
    \langle\eta(t)x(t)\rangle&=2i\delta\omega_0^2\tau\langle x(t)\rangle\cdot\frac{(2+\gamma\tau)\cos(\omega_jt)+2\omega_j\tau\sin(\omega_jt)}{(2+\gamma\tau)^2+4\omega_j^2\tau^2}\\
    \langle\phi(t)x(t)\rangle&=2i\delta\omega_0^2\tau\langle x(t)\rangle\cdot\frac{-2\omega_j\tau\cos(\omega_jt)+(2+\gamma\tau)\sin(\omega_jt)}{(2+\gamma\tau)^2+4\omega_j^2\tau^2}.
    \label{eq:phix_average}
\end{align}

Once we have the ensemble-averaged amplitude, the power is straightforward to compute from \eqref{accum_discrete}.  Taking the asymptotic ensemble average of this equation and using \eqref{shapiro_amplitude}, we find
\begin{equation}
    \langle|x(t)|^2\rangle=\frac1{\gamma\omega_0}\IM\left[\langle x(t)\rangle^*F_0e^{i\omega_0t}\right]=\frac{|F_0|^2}{\gamma^2\omega_0^2(1+\alpha)}.
\end{equation}

%%%%%%%%%%%%%%%%%%%%%%%%%%%%%%%%%%%%%%%%%%%%%%%%%%%%%%%%%%%%%%%%%%%%%%%%%%%%%%%%%%%%%%%%%%
%%%%%%%%%%%%%%%%%%%%%%%%%%%%%%%%%%%%%%%%%%%%%%%%%%%%%%%%%%%%%%%%%%%%%%%%%%%%%%%%%%%%%%%%%%
\subsection{Autocorrelation function}
\label{app:shapiro_amppower}
%%%%%%%%%%%%%%%%%%%%%%%%%%%%%%%%%%%%%%%%%%%%%%%%%%%%%%%%%%%%%%%%%%%%%%%%%%%%%%%%%%%%%%%%%%
%%%%%%%%%%%%%%%%%%%%%%%%%%%%%%%%%%%%%%%%%%%%%%%%%%%%%%%%%%%%%%%%%%%%%%%%%%%%%%%%%%%%%%%%%%

Now we compute the autocorrelation function $C_x(s)$ of the resonator, defined in \eqref{resonator_correlation}.  In order to evaluate this quantity, we require a differential equation for $\langle x(t+s)x(t)^*\rangle$.  (Throughout this section, we assume $s>0$.)  To obtain such an equation, we may evaluate \eqref{decomp_resonator} at time $t+s$, multiply by $x(t)^*$, and take the ensemble average.  As before, this will give us a differential equation in terms of other quantities, such as $\langle\eta(t+s)x(t+s)x(t)^*\rangle$.  Again we can derive a differential equation for this quantity by multiplying by $\eta(t+s)x(t)^*$ instead and applying \eqref{Shapiro Loginov}.  Utilizing the DMP property \eqref{DMP_property} and neglecting terms of the form $\langle\eta(t+s)\phi(t+s)x(t+s)x(t)^*\rangle$ to truncate this process, we arrive at the following system of equations
\begin{align}
    &L_0 \langle x(t+s)x(t)^*\rangle +2\omega_0[\langle\eta(t+s)x(t+s)x(t)^*\rangle\cos(\omega_j(t+s))\nl
    ~~~~~~~~~~~~~~~~~~~~~~~~~~+\langle\phi(t+s)x(t+s)x(t)^*\rangle\sin(\omega_j(t+s))]=F_0e^{i\omega_0(t+s)} \langle x(t)^*\rangle,\label{eq:xx_diffeq}\\
    &L_1 \langle \eta(t+s)x(t+s)x(t)^*\rangle +2\omega_0\delta\omega_0^2 \langle x(t+s)x(t)^*\rangle\cos(\omega_j (t+s))=F_0e^{i\omega_0(t+s)}\langle \eta(t+s)x(t)^*\rangle,\label{eq:etaxx_diffeq}\\
    &L_1 \langle \phi(t+s)x(t+s)x(t)^*\rangle +2\omega_0\delta\omega_0^2 \langle x(t+s)x(t)^*\rangle\sin(\omega_j (t+s))=F_0e^{i\omega_0(t+s)}\langle \phi(t+s)x(t)^*\rangle.\label{eq:phixx_diffeq}
\end{align}
Here and henceforth throughout this subsection, all derivatives in the operators $L_j$ are with respect to the variable $s$.

Using analogues of \eqref{L1cos}, along with \eqrefRange{xx_diffeq}{phixx_diffeq}, we can again apply the operator $L_1L_1$ to the LHS of \eqref{xx_diffeq} and expand it as
\begin{align}
    &L_1L_1L_0 \langle x(t+s)x(t)^*\rangle +2\omega_0L_1L_1[\langle\eta(t+s)x(t+s)x(t)^*\rangle\cos(\omega_j(t+s))+\langle\phi(t+s)x(t+s)x(t)^*\rangle\sin(\omega_j(t+s))]\\
    &=(L_1L_1L_0-4\omega_0^2\delta\omega_0^2L_1)\langle x(t+s)x(t)^*\rangle\nl
    +4i\omega_0^2\omega_jL_1[-\langle\eta(t+s)x(t+s)x(t)^*\rangle\sin(\omega_j(t+s))+\langle\phi(t+s)x(t+s)x(t)^*\rangle\cos(\omega_j(t+s))]\nl
    +2\omega_0L_1[F_0e^{i\omega_0(t+s)}\left(\langle\eta(t+s)x(t)^*\rangle\cos(\omega_j(t+s))+\langle\phi(t+s)x(t)^*\rangle\sin(\omega_j(t+s))\right)]\\
    &=(L_1L_1L_0-4\omega_0^2\delta\omega_0^2L_1)\langle x(t+s)x(t)^*\rangle\nl
    +8\omega_0^3\omega_j^2[\langle\eta(t+s)x(t+s)x(t)^*\rangle\cos(\omega_j(t+s))+\langle\phi(t+s)x(t+s)x(t)^*\rangle\sin(\omega_j(t+s))]\nl
    +4i\omega_0^2\omega_jF_0e^{i\omega_0(t+s)}\left[-\langle\eta(t+s)x(t)^*\rangle\sin(\omega_j(t+s))+\langle\phi(t+s)x(t)^*\rangle\cos(\omega_j(t+s))\right]\nl
    +2\omega_0L_1[F_0e^{i\omega_0(t+s)}\left(\langle\eta(t+s)x(t)^*\rangle\cos(\omega_j(t+s))+\langle\phi(t+s)x(t)^*\rangle\sin(\omega_j(t+s))\right)]\\
     &=(L_1L_1L_0-4\omega_0^2\delta\omega_0^2L_1-4\omega_0^2\omega_j^2L_0)\langle x(t+s)x(t)^*\rangle+4\omega_0^2\omega_j^2F_0e^{i\omega_0(t+s)}\langle x(t)^*\rangle\nl
    +4i\omega_0^2\omega_jF_0e^{i\omega_0(t+s)}\left[-\langle\eta(t+s)x(t)^*\rangle\sin(\omega_j(t+s))+\langle\phi(t+s)x(t)^*\rangle\cos(\omega_j(t+s))\right]\nl
    +2\omega_0L_1[F_0e^{i\omega_0(t+s)}\left(\langle\eta(t+s)x(t)^*\rangle\cos(\omega_j(t+s))+\langle\phi(t+s)x(t)^*\rangle\sin(\omega_j(t+s))\right)].
    \label{eq:L1L1expanded}
\end{align}
An important property of a DMP is that its higher-point correlators may be decomposed as
\begin{equation}
    \langle\eta(t_1)\eta(t_2)\cdots\eta(t_{n-1})\eta(t_n)\rangle=\langle\eta(t_1)\eta(t_2)\rangle\cdots\langle\eta(t_{n-1})\eta(t_n)\rangle
    \label{eq:DMP_npoint}
\end{equation}
when $t_1\leq t_2\leq\cdots\leq t_{n-1}\leq t_n$~\cite{bourret1973brownian}.  This implies that for any funtional $y[\eta(t)]$,
\begin{equation}
    \langle\eta(t+s)y(t)\rangle=\delta\omega_0^{-2}\langle\eta(t+s)\eta(t)\eta(t)y(t)\rangle=\delta\omega_0^{-2}\langle\eta(t+s)\eta(t)\rangle\langle\eta(t)y(t)\rangle=e^{-s/\tau}\langle\eta(t)y(t)\rangle.
\end{equation}
We can apply this property to the results in \eqref[s]{etax_average} and (\ref{eq:phix_average}) and substitute them into \eqref{L1L1expanded} to get
\begin{align}
    &(L_1L_1L_0-4\omega_0^2\delta\omega_0^2L_1-4\omega_0^2\omega_j^2L_0)\langle x(t+s)x(t)^*\rangle+4\omega_0^2\omega_j^2F_0e^{i\omega_0(t+s)}\langle x(t)^*\rangle\nl
    -8\omega_0^2\delta\omega_0^2\omega_j\tau e^{-s/
    \tau}\langle x(t)^*\rangle\cdot F_0e^{i\omega_0(t+s)}\cdot\frac{2\omega_j\tau\cos(\omega_js)+(2+\gamma\tau)\sin(\omega_js)}{(2+\gamma\tau)^2+4\omega_j^2\tau^2}\nl
    -4i\omega_0\delta\omega_0^2\tau e^{-s/\tau}\langle x(t)^*\rangle L_1\left[F_0e^{i\omega_0(t+s)}\cdot\frac{(2+\gamma\tau)\cos(\omega_js)-2\omega_j\tau\sin(\omega_js)}{(2+\gamma\tau)^2+4\omega_j^2\tau^2}\right]\\
    &=(L_1L_1L_0-4\omega_0^2\delta\omega_0^2L_1-4\omega_0^2\omega_j^2L_0)\langle x(t+s)x(t)^*\rangle+4\omega_0^2\omega_j^2F_0e^{i\omega_0(t+s)}\langle x(t)^*\rangle\nl
    -16\omega_0^2\delta\omega_0^2\omega_j\tau e^{-s/
    \tau}\langle x(t)^*\rangle\cdot F_0e^{i\omega_0(t+s)}\cdot\frac{2\omega_j\tau\cos(\omega_js)+(2+\gamma\tau)\sin(\omega_js)}{(2+\gamma\tau)^2+4\omega_j^2\tau^2}\nl
    +4\omega_0^2\delta\omega_0^2(2+\gamma\tau) e^{-s/\tau}\langle x(t)^*\rangle\cdot F_0e^{i\omega_0(t+s)}\cdot\frac{(2+\gamma\tau)\cos(\omega_js)-2\omega_j\tau\sin(\omega_js)}{(2+\gamma\tau)^2+4\omega_j^2\tau^2}.
\end{align}
In summary, we find that applying $L_1L_1$ to \eqref{xx_diffeq} gives
\begin{align}
    &(L_1L_1L_0-4\omega_0^2\delta\omega_0^2L_1-4\omega_0^2\omega_j^2L_0)\langle x(t+s)x(t)^*\rangle\nonumber\\
    &=-\omega_0^2F_0e^{i\omega_0(t+s)}\langle x(t)^*\rangle\cdot\left(\left(\gamma+\frac2\tau\right)^2+4\omega_j^2+4\delta\omega_0^2 e^{-s/\tau}\cdot\frac{\left((2+\gamma\tau)^2-8\omega_j^2\tau^2\right)\cos(\omega_js)-6\omega_j\tau(2+\gamma\tau)\sin(\omega_js)}{(2+\gamma\tau)^2+4\omega_j^2\tau^2}\right).
    \label{eq:L1L1expanded2}
\end{align}

Now \eqref{L1L1expanded2} gives us a decoupled differential equation for $C_x(s)$ [once the result for $\langle x(t)\rangle$ in \eqref{shapiro_amplitude} is substituted in].  The full solution to this equation consists of both a homogeneous contribution and an inhomogeneous contribution.  To find the former, we may take the ansatz
\begin{equation}
    \langle x(t+s)x(t)^*\rangle\sim e^{(i\omega_0+\beta)s}.
\end{equation}
Setting the RHS of \eqref{L1L1expanded2} to zero and plugging this in, we find
\begin{equation}
    \left(\gamma+\frac2\tau+2\beta\right)^2(\gamma+2\beta)+4\delta\omega_0^2\left(\gamma+\frac2\tau+2\beta\right)+4\omega_j^2(\gamma+2\beta)^2=0.
\end{equation}
The solutions to this equation are
\begin{align}
    \label{eq:beta1}
    \beta_1=&-\frac{\gamma}{2}  - \frac{1}{3\tau}\left[2 -\frac{1+i\sqrt{3}}{2}\frac{a}{\left(\sqrt{a^3+b^2}+b\right)^{1/3}}+\frac{1-i\sqrt{3}}{2}\left(\sqrt{a^3+b^2}+b\right)^{1/3}\right],\\
    \beta_2=&-\frac{\gamma}{2}  - \frac{1}{3\tau}\left[2 +\frac{a}{\left(\sqrt{a^3+b^2}+b\right)^{1/3}}-\left(\sqrt{a^3+b^2}+b\right)^{1/3}\right],\label{eq:beta2}\\
    \beta_3=&\beta_1^*,
    \label{eq:beta3}
\end{align}
where
\begin{align}
    a&=-1+3\omega_j^2\tau^2+3\delta\omega_0^2\tau^2,\\
    b&=1+9\omega_j^2\tau^2 -\frac{9}{2}\delta\omega_0^2\tau^2.
\end{align}
The homogeneous solution therefore has three contributions, with exponential dependences given by \eqrefRange{beta1}{beta3}.  In terms of the timescales introduced in the functional form in \eqref{functional_form}, these translate to $\tau_1=-1/\RE[\beta_1]$, $\omega_K=\IM[\beta_1]$, and $\tau_2=-1/\beta_2$.  In this appendix, we do not solve for the coefficients $c_1$, $c_2$, and $c_3$.

The inhomogeneous solution to \eqref{L1L1expanded2} should also contribute to $C_x(s)$.  From the RHS of \eqref{L1L1expanded2}, we can see that this inhomogeneous contribution should consist of a constant term and and two oscillatory terms with frequency $\omega_j$ and decay time $\tau$.  Note that the constant term is precisely the contribution which is subtracted off from $C_x(s)$ in the definition of the autocovariance function in \eqref{autocovariance}.  Therefore this term does not contribute to $\hat K_x(s)$.  In principle, the remaining two terms could also contribute to the functional form in \eqref{functional_form}.  The timescales for these terms are precisely the perturbative values for $\tau_1$ and $\omega_K$ predicted in \eqref{Kxhat_full}, and so we expect these terms to contribute similarly to the $\beta_1$ and $\beta_3$ homogeneous terms.  Numerically, we have re-performed the fit in \figref{autocorrelation} with additional terms corresponding to this inhomogeneous contribution and found that the fit does not significantly improve.  Therefore in \figref{autocorrelation} and \tabref{fit}, we present results with only the form shown in \eqref{functional_form}.

%%%%%%%%%%%%%%%%%%%%%%%%%%%%%%%%%%%%%%%%%%%%%%%%%%%%%%%%%%%%%%%%%%%%%%%%%%%%%%%%%%%%%%%%%%
%%%%%%%%%%%%%%%%%%%%%%%%%%%%%%%%%%%%%%%%%%%%%%%%%%%%%%%%%%%%%%%%%%%%%%%%%%%%%%%%%%%%%%%%%%
\section{Signal-to-noise ratio}
\label{app:SNR}
%%%%%%%%%%%%%%%%%%%%%%%%%%%%%%%%%%%%%%%%%%%%%%%%%%%%%%%%%%%%%%%%%%%%%%%%%%%%%%%%%%%%%%%%%%
%%%%%%%%%%%%%%%%%%%%%%%%%%%%%%%%%%%%%%%%%%%%%%%%%%%%%%%%%%%%%%%%%%%%%%%%%%%%%%%%%%%%%%%%%%

In this appendix, we derive the formula for the total SNR in \eqref{totalSNR}.  Importantly, we show that this derivation relies on the assumption that the system response $\tilde x(f)$ exhibits Gaussian behavior.  The total SNR as defined in \eqref{totalSNR} is therefore only a useful figure of merit in the frequency range near the central peak of a jittering resonator, and does not apply in the jittering-induced sidebands.

Suppose that after running our experiment for a duration $t_\mathrm{int}$, we measure a response $\tilde x(f)$.  The data $\tilde x(f)$ may have arisen purely from noise, or from a combination of a signal and noise in the system.  In order to distinguish these two scenarios, we should combine the data at all frequencies into a signal test statistic which exhibits very different distributions in these two cases.  It can be shown that the optimal bilinear test statistic [for Gaussian $\tilde x(f)$] is
\begin{equation}
    Q=\int df\,\frac{S_x^\mathrm{sig}(f)}{[S_x^\mathrm{noise}(f)]^2}\cdot|\tilde x(f)|^2.
\end{equation}
The prefactor here upweights data at frequencies with high SNR.  In the absence of a signal, the expectation of this test statistic is
\begin{equation}
    \langle Q\rangle_0=\int df\,\frac{S_x^\mathrm{sig}(f)}{[S_x^\mathrm{noise}(f)]^2}\cdot S_x^\mathrm{noise}(f)t_\mathrm{int}=t_\mathrm{int}\int df\,\frac{S_x^\mathrm{sig}(f)}{S_x^\mathrm{noise}(f)}.
\end{equation}
(Note here that we have regulated the delta function as $\delta(0)=t_\mathrm{int}$).  If $\tilde x(f)$ obeys Gaussian statistics, then we may apply Wick's theorem to compute
\begin{align}
    \langle Q^2\rangle_0&=\int dfdf'\,\frac{S_x^\mathrm{sig}(f)}{[S_x^\mathrm{noise}(f)]^2}\cdot\frac{S_x^\mathrm{sig}(f')}{[S_x^\mathrm{noise}(f')]^2}\cdot\left[S_x^\mathrm{noise}(f)S_x^\mathrm{noise}(f')t_\mathrm{int}^2+S_x^\mathrm{noise}(f)^2t_\mathrm{int}\delta(f-f')\right]\\
    &=\langle Q\rangle_0^2+t_\mathrm{int}\int df\,\left(\frac{S_x^\mathrm{sig}(f)}{S_x^\mathrm{noise}(f)}\right)^2.
    \label{eq:Q0sqr}
\end{align}
This implies that the standard deviation of $Q$ is simply $\sigma_{Q,0}=\mathrm{SNR}_\mathrm{tot}$.  On the other hand, if $\tilde x(f)$ contains both signal and noise, the expectation of the test statistic is
\begin{equation}
    \langle Q\rangle_\mathrm{sig}=\int df\,\frac{S_x^\mathrm{sig}(f)}{[S_x^\mathrm{noise}(f)]^2}\cdot\left[S_x^\mathrm{noise}(f)+S_x^\mathrm{sig}(f)\right]t_\mathrm{int}=\langle Q\rangle_0+\mathrm{SNR}_\mathrm{tot}^2.
\end{equation}
The scenarios of noise only and signal + noise are distinguishable when $\langle Q\rangle_\mathrm{sig}-\langle Q\rangle_0>\sigma_{Q,0}$.  We can readily see that this occurs precisely when $\mathrm{SNR}_\mathrm{tot}>1$, and so the total SNR as defined in \eqref{totalSNR} constitutes a relevant figure of merit when the system response is Gaussian.

Notably, our derivation of the variance of $Q$ in \eqref{Q0sqr} relied on the assumption that $\tilde x(f)$ was Gaussian.  If we attempt to apply this statistic to the sidebands of a jittering resonator's spectral response, using \eqref{nonGaussian}, we will instead find
\begin{align}
    \langle Q^2\rangle_0&=\int dfdf'\,\frac{S_x^\mathrm{sig}(f)}{[S_x^\mathrm{noise}(f)]^2}\cdot\frac{S_x^\mathrm{sig}(f')}{[S_x^\mathrm{noise}(f')]^2}\cdot2[S_x^\mathrm{noise}(f)S_x^\mathrm{noise}(f')t_\mathrm{int}^2+S_x^\mathrm{noise}(f)^2t_\mathrm{int}\delta(f-f')\nl
    ~~~~~~~~~~~~~~~~~~~~~~~~~~~~~~~~~~~~~~~~~~~~~~~~~~+S_x^\mathrm{noise}(f)^2t_\mathrm{int}\delta(f+f'-2f_0)]\\
    &=2\langle Q\rangle_0^2+4\mathrm{SNR}_\mathrm{tot}^2.
\end{align}
(under the assumption that $S_x^\mathrm{noise}(f)$ is even around $f=f_0$).  In this case, the standard deviation of $Q$ is now $\sigma_{Q,0}=\sqrt{\langle Q\rangle_0^2+4\mathrm{SNR}_\mathrm{tot}^2}$, which must be compared to $\langle Q\rangle_\mathrm{sig}-\langle Q\rangle_0=\mathrm{SNR}_\mathrm{tot}^2$.  If the SNR is large over a range $\Delta f$, then the former will be
\begin{equation}
\sigma_{Q,0}\sim t_\mathrm{int}\Delta f\cdot\frac{S_x^\mathrm{sig}(f)}{S_x^\mathrm{noise}(f)},
\end{equation}
while the latter is
\begin{equation}
\langle Q\rangle_\mathrm{sig}-\langle Q\rangle_0\sim t_\mathrm{int}\Delta f\cdot\left(\frac{S_x^\mathrm{sig}(f)}{S_x^\mathrm{noise}(f)}\right)^2.
\end{equation}
The ratio between these receives no enhancement from a longer integration time $t_\mathrm{int}$ or larger frequency range $\Delta f$.  In other words, the overall significance of this test statistic is the same as the test statistic from only a single frequency.  Therefore, incorporating data from the sidebands of a jittering resonator does not improve the sensitivity relative to the data from the central peak alone.

%%%%%%%%%%%%%%%%%%%%%%%%%%%%%%%%%%%%%%%%%%%%%%%%%%%%%%%%%%%%%%%%%%%%%%%%%%%%%%%%%%%%%%%%%%
%%%%%%%%%%%%%%%%%%%%%%%%%%%%%%%%%%%%%%%%%%%%%%%%%%%%%%%%%%%%%%%%%%%%%%%%%%%%%%%%%%%%%%%%%%
%%%%%%%%%%%%%%%%%%%%%%%%%%%%%%%%%%%%%%%%%%%%%%%%%%%%%%%%%%%%%%%%%%%%%%%%%%%%%%%%%%%%%%%%%%
%%%%%%%%%%%%%%%%%%%%%%%%%%%%%%%%%%%%%%%%%%%%%%%%%%%%%%%%%%%%%%%%%%%%%%%%%%%%%%%%%%%%%%%%%%
\end{document}